\documentclass[11pt,onecolumn, nofootinbib]{revtex4-1}
\usepackage{url,hyperref,lineno,microtype,subcaption}
\usepackage{listings}
\usepackage{bbold} 
\usepackage{amsmath} 
\usepackage{amssymb}
\usepackage[capitalise]{cleveref}

\crefname{section}{Sec.}{Secs.}
\crefname{table}{Tab.}{Tabs.}
\crefname{figure}{Fig.}{Figs.}
\crefname{equation}{Eq.}{Eqs.}
\crefname{appendix}{Appendix\ }{Appendix\ }

\usepackage[framemethod=TikZ]{mdframed}
\usepackage{xspace}
\usepackage{ulem}
\newcommand{\SARAH} {\texttt{SARAH}\xspace}

\definecolor{maroon}{cmyk}{0, 0.87, 0.68, 0.32}
\definecolor{halfgray}{gray}{0.55}
\definecolor{slha_frame}{RGB}{207, 207, 207}
\definecolor{slha_bg}{RGB}{247, 247, 247}
\definecolor{slha_red}{RGB}{186, 33, 33}
\definecolor{slha_green}{RGB}{0, 128, 0}
\definecolor{slha_cyan}{RGB}{64, 128, 128}
\definecolor{slha_purple}{RGB}{170, 34, 255}

\definecolor{mathematica_frame}{RGB}{207, 207, 207}
\definecolor{mathematica_bg}{RGB}{247, 247, 247}
\definecolor{mathematica_red}{RGB}{186, 33, 33}
\definecolor{mathematica_green}{RGB}{0, 128, 0}
\definecolor{mathematica_cyan}{RGB}{64, 128, 128}
\definecolor{mathematica_purple}{RGB}{170, 34, 255}

\lstnewenvironment{MIN}[1][]{%
\mdframed[roundcorner=5pt,backgroundcolor=mathematica_bg]
  \renewcommand{\thelstnumber}{In[\arabic{lstnumber}]}
  \lstset{language=MathIn,numbers=none,basicstyle=\ttfamily,#1,escapeinside=||}%
}{%
\endmdframed
}

\lstnewenvironment{MOUT}[1][]{%
\mdframed[roundcorner=5pt,backgroundcolor=mathematica_bg]
  \renewcommand{\thelstnumber}{Out[\arabic{lstnumber}]}
  \lstset{language=MathOut,numbers=left,basicstyle=\ttfamily,#1}%
}{%
\endmdframed
}

\usepackage{listings}
\lstdefinelanguage{Fortran90}{
    morekeywords={Real,Complex,Intent},%
    emph={End,Subroutine,dp,in,Function,Implicit,None},%
    emphstyle={\color{mathematica_purple}},    
    sensitive=true,%
    morecomment=[l]\#,%
    morestring=[b]',%
    morestring=[b]",%
    morestring=[s]{'''}{'''},
    morestring=[s]{"""}{"""},
    morestring=[s]{r'}{'},
    morestring=[s]{r"}{"},%
    morestring=[s]{r'''}{'''},%
    morestring=[s]{r"""}{"""},%
    morestring=[s]{u'}{'},
    morestring=[s]{u"}{"},%
    morestring=[s]{u'''}{'''},%
    morestring=[s]{u"""}{"""},%
    identifierstyle=\color{black}\ttfamily,
    commentstyle=\color{slha_cyan}\ttfamily,
    stringstyle=\color{slha_red}\ttfamily,
    keepspaces=true,
    showspaces=false,
    showstringspaces=false,
    rulecolor=\color{slha_frame},
    frame=true,
    frameround={t}{t}{t}{t},
    framexleftmargin=6mm,
    numbers=left,
    numberstyle=\tiny\color{halfgray},
    backgroundcolor=\color{slha_bg},
    basicstyle=\footnotesize,
    keywordstyle=\color{slha_green}\ttfamily,
    aboveskip=1.2em,
    belowskip=1.2em,
}

\lstdefinelanguage{SLHA}{
    morekeywords={block,Block,BLOCK,decay,Decay,DECAY},%
    sensitive=true,%
    morecomment=[l]\#,%
    morestring=[b]',%
    morestring=[b]",%
    morestring=[s]{'''}{'''},
    morestring=[s]{"""}{"""},
    morestring=[s]{r'}{'},
    morestring=[s]{r"}{"},%
    morestring=[s]{r'''}{'''},%
    morestring=[s]{r"""}{"""},%
    morestring=[s]{u'}{'},
    morestring=[s]{u"}{"},%
    morestring=[s]{u'''}{'''},%
    morestring=[s]{u"""}{"""},%
    identifierstyle=\color{black}\ttfamily,
    commentstyle=\color{slha_cyan}\ttfamily,
    stringstyle=\color{slha_red}\ttfamily,
    keepspaces=true,
    showspaces=false,
    showstringspaces=false,
    rulecolor=\color{slha_frame},
    frame=true,
    frameround={t}{t}{t}{t},
    framexleftmargin=6mm,
    numbers=left,
    numberstyle=\tiny\color{halfgray},
    backgroundcolor=\color{slha_bg},
    basicstyle=\footnotesize,
    keywordstyle=\color{slha_green}\ttfamily,
    aboveskip=1.2em,
    belowskip=1.2em,
}

\lstdefinestyle{terminal} {
  morekeywords={cp,-r,make,cd},
  numbers=left, 
  stepnumber=1, 
  numberstyle=\tiny\color{halfgray}, 
  numbersep=10pt, 
  backgroundcolor=\color{black}, 
  basicstyle=\color{white}\ttfamily,
  stringstyle=\color{white}\ttfamily,
  keywordstyle=\color{white}\ttfamily\bfseries
 }

\lstdefinelanguage{MathIn}{
    morekeywords={Simplify,Eigenvalues,epsUV,Delta,UVscaleQ},%
    emph={Start,InitUnitarity,GetScatteringDiagrams,BuildScatteringMatrix,MakeSPheno,InitMatching,EFTcoupLO,EFTcoupNLO},%
    emphstyle={\color{mathematica_purple}},
    sensitive=true,%
    morecomment=[l]\%,%
    morestring=[b]',%
    morestring=[b]",%
    morestring=[s]{'''}{'''},
    morestring=[s]{"""}{"""},
    morestring=[s]{r'}{'},
    morestring=[s]{r"}{"},%
    morestring=[s]{r'''}{'''},%
    morestring=[s]{r"""}{"""},%
    morestring=[s]{u'}{'},
    morestring=[s]{u"}{"},%
    morestring=[s]{u'''}{'''},%
    morestring=[s]{u"""}{"""},%
    identifierstyle=\color{black}\ttfamily,
    commentstyle=\color{mathematica_cyan}\ttfamily,
    stringstyle=\color{mathematica_red}\ttfamily,
    keepspaces=true,
    showspaces=false,
    showstringspaces=false,
    rulecolor=\color{mathematica_frame},
    frame=none,
    numbers=left,
    numberstyle=\tiny\color{halfgray},
    basicstyle=\footnotesize,
    keywordstyle=\color{mathematica_green}\ttfamily,
    aboveskip=0.2em,
    belowskip=0.2em
}

\lstdefinelanguage{MathOut}{
    morekeywords={Simplify,Eigenvalues},%
    sensitive=true,%
    morecomment=[l]\%,%
    morestring=[b]',%
    morestring=[b]",%
    morestring=[s]{'''}{'''},
    morestring=[s]{"""}{"""},
    morestring=[s]{r'}{'},
    morestring=[s]{r"}{"},%
    morestring=[s]{r'''}{'''},%
    morestring=[s]{r"""}{"""},%
    morestring=[s]{u'}{'},
    morestring=[s]{u"}{"},%
    morestring=[s]{u'''}{'''},%
    morestring=[s]{u"""}{"""},%
    identifierstyle=\color{black}\ttfamily,
    commentstyle=\color{mathematica_cyan}\ttfamily,
    stringstyle=\color{mathematica_red}\ttfamily,
    keepspaces=true,
    showspaces=false,
    showstringspaces=false,
    rulecolor=\color{mathematica_frame},
    frame=none,
    frameround={t}{t}{t}{t},
    framexleftmargin=10mm,
    numbers=left,
    numberstyle=\tiny\color{halfgray},
    basicstyle=\footnotesize,
    keywordstyle=\color{mathematica_green}\ttfamily,
    aboveskip=0.2em,
    belowskip=0.2em,
}

\let\origthelstnumber\thelstnumber
\makeatletter
\newcommand*\Suppressnumber{%
  \lst@AddToHook{OnNewLine}{%
    \let\thelstnumber\relax%
     \advance\c@lstnumber-\@ne\relax%
    }%
}

\newcommand*\Reactivatenumber{%
  \lst@AddToHook{OnNewLine}{%
   \let\thelstnumber\origthelstnumber%
   \advance\c@lstnumber\@ne\relax}%
}

\newcommand{\nn}{\nonumber}
\newcommand{\SPheno} {\texttt{SPheno}\xspace}

\def\lsim{\raise0.3ex\hbox{$\;<$\kern-0.75em\raise-1.1ex\hbox{$\sim\;$}}}
\def\gsim{\raise0.3ex\hbox{$\;>$\kern-0.75em\raise-1.1ex\hbox{$\sim\;$}}}

\begin{document}

 \author{T. Faber}
 \email[E-mail: ]{thomas.faber@physik.uni-wuerzburg.de}
 \author{Y. Liu}
 \email[E-mail: ]{yang.liu@uni-wuerzburg.de}
 \author{W. Porod}
 \email[E-mail: ]{porod@physik.uni-wuerzburg.de}
 \affiliation{
 Institut f\"ur Theoretische Physik und Astrophysik, Universit\"at W\"urzburg, Germany
 }

 \author{M. Hudec}
 \email[E-mail: ]{hudec@ipnp.troja.mff.cuni.cz}
 \author{M. Malinsk\'y}
 \email[E-mail: ]{malinsky@ipnp.troja.mff.cuni.cz}
 \affiliation{
  Institute for Particle and Nuclear Physics, Charles University, Czech Republic
  }

 \author{F. Staub}
 \email[E-mail: ]{florian.staub@kit.edu}
 \affiliation{
  Institute for Nuclear Physics, Karlsruhe Institute of Technology, Germany
 }

 \author{H. Kole\v{s}ov\'a}
 \email[E-mail: ]{helena.kolesova@uis.no}
 \affiliation{
  University of Stavanger, Norway
 }
\title{Collider phenomenology of a unified leptoquark model
}


\begin{abstract}
We demonstrate that in a recently proposed unified leptoquark model
based on the gauge group $SU(4)_C\times SU(2)_L \times U(1)_R$
significant deviations from the Standard Model values of $R_K$ and $R_{K^*}$ can be accommodated without any need of extra heavy
fermions. Low energy data, in particular lepton flavour
violating $\mu$ decays and $K_L\to e \mu$,  severely constrain the 
available parameter space. We show that in the allowed part of the parameter
space (i) some of the lepton-flavour-violating tau decay branching ratios
are predicted to be close to their current experimental limits. (ii) The underlying  scalar leptoquarks can be probed at the LHC via their dominant decay modes into tau-leptons and electrons
and the third generation quarks. (iii) The constraints from meson oscillations imply that the masses of scalar gluons, another pair of coloured multiplets around, have to be bigger than around 15 TeV and, thus,
they can be probed only at a future 100 TeV collider. In both neutral and charged variants, these scalars decay predominantly into third generation quarks,
with up to $O$(10\%) branching ratios into family-mixed final states. Besides that, we comment on the phenomenology of the scalar
gluons in the current scenarios in the case that the $B$-decay anomalies eventually disappear. 
\end{abstract}
\maketitle

\section{Introduction}
The latest results of the LHC clearly show that the 
Standard Model (SM) continues to be a remarkably successful description of 
nature. So far, only a  handful of experimental observations
show deviations from its predictions.  At the moment, exciting
direct hints of physics beyond the SM  are the recently observed
anomalies in $B$-meson decays \cite{Matyja:2007kt,Bozek:2010xy,Aaij:2014ora,Huschle:2015rga,Aaij:2015yra,Aaij:2017vbb}, which suggest 
lepton flavour universality violation (LFUV) in the 
ratios $R_{K^{(*)}}=\Gamma(\bar B \to \bar K^{(*)} \mu^+ \mu^-)/
                   \Gamma(\bar B \to \bar K^{(*)} e^+ e^-)$   and
$R_{D^{(*)}} = \Gamma(\bar B \to D^{(*)} \tau \bar \nu)/
                   \Gamma(\bar B \to D^{(*)} l \bar \nu)$,
        $(l=e,\mu)$,
with, e.g. \cite{Aaij:2019wad},
\begin{align}
R_K =
{0.846\,^{+\,0.060}_{-\,0.054}\text{ (stat)}\,^{+\,0.016}_{-\,0.014}}\text{ (syst)}\,.
\end{align}

Assuming that these anomalies are not a result of experimental systematics, 
they can be accounted for by leptoquarks (LQs) of various kinds
\cite{Bauer:2015knc,Bauer:2015boy,Chao:2015nac,Murphy:2015kag,Alonso:2015sja,Calibbi:2015kma,Fajfer:2015ycq,Hiller:2016kry,%
Deppisch:2016qqd,Bhattacharya:2016mcc,Barbieri:2017tuq,%
Buttazzo:2017ixm,Alok:2017sui,Kumar:2018kmr,Biswas:2018iak}.
However, building models in which these fields emerge from the extended gauge symmetries is generally rather challenging, especially in the light of very stringent 
constraints on lepton flavour violation (LFV) from various
experimental searches, see e.g.~\cite{Crivellin:2017zlb,Crivellin:2017dsk}.

Several attempts to build UV complete SM extensions of this kind already exist in the literature
\cite{DiLuzio:2017vat,Calibbi:2017qbu,Bordone:2017bld,Barbieri:2017tuq,Dorsner:2017ufx,Assad:2017iib%
,Blanke:2018sro,Greljo:2018tuh,Bordone:2018nbg,Matsuzaki:2018jui,Heeck:2018ntp,Balaji:2018zna,Fornal:2018dqn}.
Most of them aim
at getting the vector leptoquark $U_1$  (cf.~\cite{Dorsner:2016wpm}), transforming as $(3,1,+2/3)$ under the SM gauge group
$G_\mathrm{SM}=SU(3)_c\times SU(2)_L \times U(1)_Y$, sufficiently light as it is
an excellent candidate to explain the anomalies. It emerges naturally
from the breaking of $SU(4)_C$ to $SU(3)_c$ which fixes the properties
of $U_1$ up to effects from generation mixing of the fermions to which it couples.
However, in most of these works the details of the scalar sector, e.g.\ the masses and couplings of the scalars, have been ignored.

To this end, we have recently   \cite{Faber:2018qon}  presented a detailed analysis of a model \cite{Smirnov:1995jq,Perez:2013osa} featuring an
$SU(4)_C\times SU(2)_L\times U(1)_R$  gauge symmetry and a minimal fermionic content 
in which 
the freedom in the scalar sector in principle allows for an explanation
of $R_{K^{(*)}}$ 
even with the $U_1$ mass above 1000 TeV (as required by the stringent $K_L$ decay constraints).
In doing so, we have used the $SO(10)$--inspired simplifying assumption
that all Yukawa couplings are symmetric in the flavour indices (in the defining basis).
This hypothesis, however, turned out to be too restrictive as it does not 
 resolve the  tension between $R_K$ and the bounds on $K_L\rightarrow e\mu$.

In this paper we show that when releasing the symmetry conditions on the Yukawa matrices significant deviations from the SM values of $R_{K^{(*)}}$ in the direction indicated by the experiment can be accommodated 
without violating any other experimental bound. In the scheme under consideration, the scalar leptoquark $R_2$ couples dominantly to the electrons; needless to say, in such a case one cannot address the discrepancies observed in the 
angular distributions of the decay $B\to K^* \mu^+ \mu^-$~\cite{Aaij:2015oid}.
On the other hand, the allowed parameter space is quite restricted 
which implies that the properties of
the additional scalars are fixed to a high degree. Consequently,
this leads to rather specific predictions for LHC searches.

The paper is organized as follows: in \cref{sec:model} we summarize
the main features of the model with a particular focus on its aspects
relevant for the $B$-physics anomalies. In \cref{sec:lowenergy} we
discuss various constraints stemming from the low energy data and
their consequences for the properties of the new scalars. This is
followed by a discussion of the resulting collider phenomenology 
in \cref{sec:collider}. A brief summary is given in \cref{sec:conclusions}.

For our investigation we used  the \SARAH package
\cite{Staub:2008uz,Staub:2009bi,Staub:2010jh,Staub:2012pb,Staub:2013tta} which
needed to be extended considerably. We present this extension
in Appendix \ref{app:SARAH}. 
For the numerical calculations we used the generated model files to produce a spectrum generator based on
\SPheno~\cite{Porod:2003um,Porod:2011nf}. For the calculation of
cross section at hadron colliders we have used the \SARAH-generated
interface to {\tt MadGraph\_aMC@NLO} \cite{Alwall:2011uj,Alwall:2014bza}.

\section{Model aspects}\label{sec:model}

We briefly summarize here the main features of the model
that are important for the subsequent discussion. For further
details we refer to 
refs.~\cite{Smirnov:1995jq,Perez:2013osa,Faber:2018qon}.
The Model is based on the gauge group  
$G=SU(4)_C \otimes SU(2)_L\otimes U(1)_R$ where the SM $SU(3)_c$ 
emerges as part of the $SU(4)_C$ factor.  In this class of models, leptons (including 
the right-handed neutrinos) are unified with quarks in representations of $G$ as 
summarized in \cref{tab:Fields}. The sub-eV neutrino masses and the observed leptonic mixing pattern are accommodated via an inverse seesaw
mechanism \cite{Mohapatra:1986bd} by adding  3 extra generations of a gauge-singlet fermion 
$N_L$  to the original model of ref.~\cite{Smirnov:1995jq} as proposed in 
\cite{Perez:2013osa}.
The inverse seesaw is the only source of the lepton number violation while $B$
remains a good symmetry to all orders in perturbation theory~\cite{Faber:2018qon}. 

\begin{table}[t]
    \centering
    \begin{tabular}{l|l}
\textbf{Fermions}  & \textbf{Scalars} \\ \hline
$F_L \,_{\, (4,2,0)}= \begin{pmatrix}Q\\L\end{pmatrix}$ 
&    $\chi_{\, (4,1,+1/2)} =
    \begin{pmatrix}  
    {\bar{S}_1^\dagger}{}_{(3,1,+2/3)} 
    \\ {\chi^0}_{\,(1,1,0)} 
    \end{pmatrix}$     
\\
${f^u_R}_{\,(4,1,+1/2)}=\begin{pmatrix}u_R\\ \nu_R \end{pmatrix}$ &
$H_{\,(1,2,+1/2)}$ 
\\ 
${f^d_R}_{\,(4,1,-1/2)}=\begin{pmatrix}d_R \\ e_R\end{pmatrix}$ &
$ \Phi_{\, (15,2,+1/2)}=
    \begin{pmatrix} 
    G_{\,(8,2,+1/2)} + \frac{\mathbb 1}{\sqrt{12}}H_2 
    & {R_2}_{\,(3,2,+7/6)} 
    \\ {{{\tilde{R}}_2^\dagger}}{}_{\,( \bar 3 , 2, -1/6)}
    &\frac{-3}{\sqrt{12}}{H_2}_{\,(1,2,+1/2)}   
    \end{pmatrix} $
\\             
$N_L \,_{\,(1,1,0)}$ & 
\end{tabular}
    \caption{Fermion and scalar content of the model at the 
    $G=SU(4)_C \otimes SU(2)_L\otimes U(1)_R$ and $G_\mathrm{SM}$ 
    levels, respectively.
    }
    \label{tab:Fields}
\end{table}

\subsection{Symmetry breaking and scalar sector}

The scalar sector consists of three irreducible representations of $G$ as given in \cref{tab:Fields}. At the level of $G_\mathrm{SM}$, the colorless part includes a complex singlet $\chi^0$ and two Higgs doublets $H$ and $H_2$. The gauge symmetry is broken by their vacuum expectation values (VEVs) in two steps
\begin{align}
    G \xrightarrow{\langle\chi^0\rangle, \langle H_2 \rangle} G_\mathrm{SM} 
    \xrightarrow{\langle H\rangle, \langle H_2 \rangle} 
    SU(3)_c\otimes U(1)_Q\,.
\end{align}
We parametrize the VEVs as
\begin{align}
    \langle \chi^0 \rangle &=\frac{v_\chi}{\sqrt{2}},    &
    \langle H \rangle &=\frac{\sin\beta}{\sqrt{2}} \begin{bmatrix}0\\ v_\mathrm{ew}   \end{bmatrix},    &
   \langle H_2\rangle &=\frac{\cos\beta}{\sqrt{2}} \begin{bmatrix}0\\ v_\mathrm{ew}   \end{bmatrix},
\end{align}
where the square brackets denote the $SU(2)_L$ doublet structure, 
$v_\mathrm{ew} = 246$~GeV  
and $v_\chi \approx 1000$~TeV. The latter is chosen such that  the vector leptoquark mass
is consistent with the stringent bound\footnote{This bound can be actually lowered by more than an order of magnitude if one maximally exploits  the freedom in the associated unitary charged-current interaction matrix \cite{Smirnov:2018ske};
however, in the current study we need to save this freedom for configuring the \emph{scalar} leptoquark interactions.}
set by the non-observation of $K_L\to e \mu$.

As usual in the two-Higgs-doublet models (2HDM), it is convenient to rotate the $SU(2)$ doublets via
\begin{equation}
\begin{pmatrix}  \hat H \\ h \end{pmatrix}
= \begin{pmatrix} \cos\beta & -\sin\beta \\ \sin \beta & \cos \beta \end{pmatrix}
\begin{pmatrix}
H \\ H_2
\end{pmatrix}\,,
\label{eq:HiggsRotation}
\end{equation}
where $h$ accommodates the entire electroweak VEV and contains also the would-be Nambu-Goldstone bosons to be eaten by $W^\pm$ and $Z$, 
whereas $\hat H$ is a second Higgs doublet which does not participate
at the electroweak symmetry breaking.
One can follow the analogy with the 2HDMs one step further. In particular, the physical component of the $h$ field defined by transformation \eqref{eq:HiggsRotation}
corresponds almost exactly to the SM Higgs because the current setting may be viewed as the 2HDM in the decoupling regime as the $\hat H$ mass is expected to be pushed up to the $SU(4)_C$
breaking scale $v_\chi$. Furthermore, the admixture of $\chi^0$ in the physical Higgs is also suppressed by $v_{\rm ew}/v_\chi$. All this can be readily verified by the analysis of the most general renormalizable scalar potential\footnote{The completeness of formula \eqref{eq:Vscalar} can be readily verified by computer codes such as {\tt Sym2Int} \cite{Fonseca:2017lem}.
}
\begin{align}
V &= \mu_H^2 |H|^2 + \mu_\chi^2 |\chi|^2 + \mu_\Phi^2 {\rm{Tr}} (|\Phi|^2) + \lambda_1 |H|^2 |\chi|^2
+ \lambda_2 |H|^2  {\rm{Tr}} (|\Phi|^2)
+ \lambda_3 |\chi|^2  {\rm{Tr}} (|\Phi|^2)
\nn\\ & 
+ ( \lambda_4 H^\dagger_i \chi^\dagger \Phi^i \chi + {\rm{h.c.}} )
+ \lambda_5 H^\dagger_i {\rm{Tr}} (\Phi^\dagger_j \Phi^i)  H^j
+ \lambda_6 \chi^\dagger \Phi^i \Phi^\dagger_i \chi
+ \lambda_7 |H|^4 + \lambda_8 |\chi|^4
+ \lambda_9 {\rm{Tr}} (|\Phi|^4)
\nn\\&
+  \lambda_{10} ({\rm{Tr}} |\Phi|^2)^2
+\Big( \lambda_{11} H^\dagger_i \;\mathrm{Tr}( \Phi^{i} \, \Phi^{j})  H^\dagger_j
+\lambda_{12} H^\dagger_i \;\mathrm{Tr}(\Phi^{i}\, \Phi^{j} \,\Phi^{\dagger }_{j})
+
\lambda_{13} H^\dagger_i \;\mathrm{Tr}( \Phi^{i} \,\Phi^{\dagger }_{j} \,  \Phi^{j})  + \mathrm{h.c.}\Big) \nn\\
&
+\lambda_{14} \chi^\dagger  |\Phi|^2  \chi
+\lambda_{15} \mathrm{Tr}( \Phi^{\dagger}_{i}\, \Phi^{j}_{}\, \Phi^{\dagger}_{j}\, \Phi^{i}_{})
+\lambda_{16} \mathrm{Tr}( \Phi^{\dagger}_{i}\, \Phi^{j}_{})\,
\mathrm{Tr}( \Phi^{\dagger}_{j}\, \Phi^{i}_{})
+\lambda_{17} \mathrm{Tr}( \Phi^{\dagger}_{i}\, \Phi^{\dagger}_{j})\, \mathrm{Tr}( \Phi^{i}_{}\, \Phi^{j}_{}) \nn\\
& +\lambda_{18} \mathrm{Tr}( \Phi^{\dagger}_{i}\, \Phi^{\dagger}_{j}\, \Phi^{i}_{}\, \Phi^{j}_{})
+\lambda_{19} \mathrm{Tr}( \Phi^{\dagger}_{i}\, \Phi^{\dagger}_{j}\, \Phi^{j}_{}\, \Phi^{i}_{}),
\label{eq:Vscalar}
\end{align}
where $|H|^2=H^\dagger_i H^i$, $|\chi|^2 = \chi^\dagger \chi$, $|\Phi|^2 = \Phi^\dagger_i \Phi^i$ with $i$ and $j$ denoting the $SU(2)_L$ indices; matrix notation has been used to capture the $SU(4)_C$ structure and the
traces run only over the $SU(4)_C$ indices.

The coloured scalar degrees of freedom are the $\bar{S}_1$ field originating from $\chi$ which dominates the Goldstone mode associated with the vector leptoquark, an $SU(2)_L$ doublet $G$  of charged and neutral scalar gluons and two other leptoquark doublets $R_2$ and $\tilde{R}_2$, all of them stemming from $\Phi$.

Although we have chosen $v_\chi$ so large that the effects of the extra vector bosons (the $Z'$ and the vector leptoquark $U_1$) are completely negligible, the model allows for a certain part of the scalar spectrum being much lighter. This can be easily seen by neglecting for the moment the effects
of the $SU(2)_L$ breaking VEVs in the masses 
of the different components of the $\Phi$-field:\footnote{Needless to say, the weak isospin mass splitting for a heavy doublet $X$ is only of the order $\delta m_X = O(v_\mathrm{ew}^2 / m_X)\lesssim 10$ GeV.}

\begin{align}
m^2_G&= \left(\frac{\sqrt{3} \lambda _4 }{4 }\tan\beta
-\frac{3}{8} \left(\lambda _6+\lambda _{14}\right)\right)v_{\chi }^2\,,\label{eq:MassOfG}\\
m^2_{R_2}&= \left(\frac{\sqrt{3} \lambda _4 }{4 } \tan\beta
	+\frac{\lambda _{14}-3 \lambda _6}{8}\right)v_{\chi }^2\,,	\label{eq:MassOfR2}\\
m^2_{\tilde{R}_2}&= \left(\frac{\sqrt{3} \lambda _4 }{4 } \tan\beta
	+\frac{\lambda _6-3\lambda _{14}}{8}\right)v_{\chi }^2\,,	\label{eq:MassOftildeR2}\\
m^2_{\hat{H}} &= \frac{\sqrt{3} \lambda_4}{2 \sin(2\beta)}  v_{\chi }^2\,,
\label{eq:MassOfH2} 
\end{align}
where $\mu_\Phi^2$ has been eliminated using the minimization conditions for the potential. This 
yields  an approximate tree-level sum rule \cite{Faber:2018qon} 
\begin{equation}
m^{2}_{G}+2 m^{2}_{\hat{H}}\sin^{2}\beta =\frac{3}{2}(m^{2}_{R_{2}}+m^{2}_{\tilde{R}_{2}})\,.
\label{massSumRule}
\end{equation}
It is well known that, unlike $\tilde{R}_2$, the $R_2$ leptoquark has the potential to simultaneously accommodate $R_K<1$ and $R_{K^*}<1$~\cite{DAmico:2017mtc}. From \cref{massSumRule} one can see that $R_2$ can be in the TeV range even in the case of a rather large
$v_\chi$ if there is an appropriate fine tuning between $\lambda_4 \tan\beta$,
$\lambda_6$ and $\lambda_{14}$ such that the entire bracket in Eq.~(\ref{eq:MassOfR2}) is suppressed to the ${\cal O}(10^{-6})$ level. 
 
Assuming for the moment 
 that $\lambda_4$ is at least of the order of $10^{-2}$, one can see from \cref{massSumRule} that relatively light scalar gluons  are possible in scenarios where $R_2$ is light and
 $\tilde R_2$ heavy. 
 We will thus investigate also such scenarios. In principle
 also $\lambda_4$ could be smaller yielding somewhat lighter $\hat H$ and $\tilde R_2$
states. However, the contribution of $\tilde R_2$ to lepton flavour violating
 observables implies that the masses should be 
in the multi-TeV range.
For completeness, we note that the large number
of parameters allows to obtain easily a SM-like
Higgs boson with $m_{h^0}=125$ GeV. 
Since the purely scalar interaction vertices play a negligible role in the phenomenology under consideration we shall not specify the particular choices of $\lambda$'s there.

\subsection{Fermionic sector}\label{sec:fermions}
The fermion masses are generated by the following Lagrangian: 
\begin{align}
\label{eq:YukawaL}
-\mathcal L_Y &=
    \overline{f^u_R} Y_1 H F_L 
    + \overline{f^u_R} Y_2 \Phi F_L 
    + \overline{f^d_R} Y_3 H^\dagger F_L 
    + \overline{f^d_R} Y_4 \Phi^\dagger F_L 
    + \overline{f^u_R} Y_5 \chi N_L 
    + \frac 1 2 N_L^T \mathcal C \mu N_L + \text{h.c.}\,,
\end{align}
where $Y_i$ are matrices of Yukawa couplings
and $\mu$ is a Majorana mass matrix.
Without loss of generality, we work in a basis where the charged-lepton mass matrix is flavour-diagonal. The up- and down-type quarks in the mass basis are given by $\hat q_L = V_q q_L$ and $\hat q_R=  U_q {q_R} $ for $q=u,d$, with the four arbitrary unitary matrices in the flavour space being constrained by $V_{\rm CKM} = V_u {V_d}^\dagger$.

Two of the Yukawa matrices above are strongly related to the masses of down-type quarks and charged leptons, namely
\begin{align}
U_d^\dagger \hat{M}_d V_d = &\left(\frac{\sin\beta}{\sqrt{2}} Y_3 + \frac{\cos\beta}{2 \sqrt{6}} Y_4 \right) v_\mathrm{ew}, \label{eq:massD}
\\
\label{eq:massL}
\hat{M}_e= & \left(\frac{\sin\beta}{\sqrt{2}} Y_3 - \frac{3 \cos\beta}{2\sqrt{6}} Y_4 \right) v_\mathrm{ew},
\end{align}where $\hat M_{u,d,e}$ are diagonal matrices of the corresponding fermion masses. 

The Yukawa interactions of the LQs and scalar gluons are encoded solely in $Y_2$ and $Y_4$. Eqs.~\eqref{eq:massD} and \eqref{eq:massL} determine $Y_4$ up to the two rotation matrices. On the other hand, due to the extended neutrino sector, the other important matrix $Y_2$, as well as $Y_5$, can be chosen essentially arbitrarily. Indeed, the measured up-type quark masses satisfying
\begin{align}
U_u^\dagger \hat{M}_u V_\mathrm{CKM} V_d   
= \left(\frac{\sin\beta}{\sqrt{2}} Y_1 + \frac{\cos\beta}{2 \sqrt{6}} Y_2 \right) v_\mathrm{ew}
\end{align}
can be always attained by a suitable choice of $Y_1$.
The light Majorana neutrino mass matrix, from which the neutrino masses and PMNS matrix follow, can be then obtained via a proper choice of the Majorana mass matrix $\mu$. 

While both $Y_2$ and $Y_4$ do in general contribute to various lepton-flavour violating processes, only $Y_4$ is relevant for a tree-level explanation of the $R_{K^{(*)}}$ anomalies. 
For simplicity, we will assume that all elements of $Y_2$ are negligibly small (see \cref{tab:parameters} for our particular choice of the \SPheno input). As will be clarified in Sect.~\ref{sec:lowenergy}, the main reason for this is the need to satisfy the very stringent limits on the LFV muon decays ($\mu \to e \gamma$
and $\mu\to 3 e$, see \cite{Faber:2018qon}) as well as other constraints such as those coming from $\tau\to e\pi^0$ etc.; in this respect, the situation with $Y_4$ alone is much ``safer" than that of any significant interplay among the two. It is also worth noting that in this case the specific form of $U_u$ is not important as vertices  where $Y_4$ appears do not contain right-handed up quarks. The only other sector it affects is neutrinos; there it enters together with $Y_5$ which, however, entertains a lot of freedom anyway. For definiteness, in what follows
we set $U_u=U_d V_\mathrm{CKM}^\dagger$. Note also that, unless specified otherwise, the results below (especially those of Sect.~\ref{subs:colliderAnomalies}) are quite robust with respect to invoking small but non-zero~$Y_2$.        

\section{Rare lepton and meson decays}
\label{sec:lowenergy}

As explained in Section~\ref{sec:model}, we assume that the only relatively light BSM field around is the leptoquark doublet $R_2$ and all other heavy fields are effectively decoupled. Cases with other relatively light extra multiplets are discussed at the end of this section. As outlined in the previous section we assume that $Y_2$ is small and, thus, the only relevant BSM matter interactions are those following from the term proportional to $Y_4$. For the $R_2$ leptoquark, these read
\begin{align}
\mathcal{L}_{R_2} 
=\overline{\hat d_L}\,  \hat{Y}_4^{de} \, \hat e_R \, R_2^{+2/3}
+ \overline{\hat u_L}\, V_\mathrm{CKM} \hat{Y}_4^{de} \, \hat e_R \, R_2^{+5/3} 
+ \text{h.c.}\,,
\label{L4}
\end{align}	
with the relevant Yukawa matrices parametrized as
\begin{align}
\hat Y_4^{de}=
\begin{pmatrix}
y_{de}&y_{d\mu}&y_{d\tau}\\
y_{se}&y_{s\mu}&y_{s\tau}\\
y_{be}&y_{b\mu}&y_{b\tau}
\end{pmatrix}
\,,\qquad
\hat V_\mathrm{CKM} \hat Y_4^{de}=
\begin{pmatrix}
y_{ue}&y_{u\mu}&y_{u\tau}\\
y_{ce}&y_{c\mu}&y_{c\tau}\\
y_{te}&y_{t\mu}&y_{t\tau}
\end{pmatrix}\,.
\label{Yde}
\end{align}

\subsection[Constraints on the Y4 structure]{Constraints on the $\hat{Y}_4$ structure}
Without referring to the specific pattern of the  matrix above imposed by the extended symmetry of the model (cf. Eqs.~\eqref{eq:massD} and \eqref{eq:massL}) several simple but important observations can be made. 

First, the  interactions in \cref{L4}
involve the right-handed leptons. In view of $R_K$, this implies that 
the corresponding tree-level contributions to $C_9$ and $C_{10}$ 
(entering at the scale where the leptoquarks are integrated out) have not only
the same magnitude but also the same sign. Since the SM contribution yields $C_9^\mathrm{SM} \approx - C_{10}^\mathrm{SM}$ there is only a very small interference between the NP and the SM contributions in the $b\to s l^+ l^-$ amplitudes. 
Notice that there are ways to circumvent this feature by making the loop contributions dominant, see \cite{Becirevic2017, Fajfer:2018bfj}; this, however, is not applicable in the current scenario. 

Second, interactions in \cref{L4} generally induce new sources of LFUV whenever two columns of $ \hat{Y}_4^{de}$ differ. In this respect,  
$R_K<1$ can be achieved if and only if the LQs couple more to the electrons than to the muons \cite{DAmico:2017mtc}, i.e., when $|y_{se} y_{be}|>|y_{s\mu}y_{b\mu}|$.

The third point is that the interactions in \cref{L4} mediate lepton flavour violating (LFV) processes whenever there are nonzero entries of $\hat Y_4^{de}$ in two different columns. 
For example, very stringent constraints arise from the experimental limits on $\mathrm{BR}(K^0_L \rightarrow e^\pm \mu^\mp )\propto|y_{se} y_{d\mu}^* + y_{de} y_{s\mu}^*|^2$  or from $\mu\to e \gamma$ whose amplitudes are given by linear combinations of $y_{qe} y_{{q'}\mu}^* $. To this end, it is clear that all the muon number violating processes mediated by $R_2$  will be suppressed if
\begin{align}
y_{d\mu}=y_{s\mu}=y_{b\mu}=0
\label{muonOut}
\end{align}approximately holds.

As indicated earlier, $\hat Y_4^{de}$ cannot be chosen arbitrarily in our model as it is a subject of the extended symmetry constraints. In particular, applying the flavour rotations in \cref{eq:YukawaL} and using  relations (\ref{eq:massD}) and (\ref{eq:massL}) one obtains the following pattern \cite{Popov:2005wz}:
\begin{align}
\hat Y_4^{de}=
 \frac{\sqrt{3/2}}{v_\mathrm{ew} \cos\beta} 
\left(	  \hat M_{d} U_d - V_d 	\hat M_e \right)\,,
\label{YdeSU4}
\end{align}
with $U_d$ and $V_d$ being arbitrary unitary matrices.
The question now is whether this structure is compatible with $R_K$ significantly smaller than 1 and a suppressed LFV pattern. 

In Ref.~\cite{Faber:2018qon}, this model was studied under an extra  $SO(10)$ inspired assumption $V_d = U^*_d$ and with all possible phases neglected in
a second step.
In such a case, the interaction matrix in \cref{YdeSU4} simplifies to
\begin{align}
\hat Y_4 ^{de}
=\sqrt{\frac{3}{2}}\frac{\sqrt{1+\tan ^2\beta}}{v_\mathrm{ew}}
\begin{pmatrix}
 V_{11} \left(m_{\text{d}}\! - \! m_{\text{e}}\right) 
& V_{12}\left(m_{\text{d}}\! - \! m_{\mu }\right)
& V_{13}  \left(m_{\text{d}}\! - \! m_{\tau }\right)
\\
V_{21} \left(m_{\text{s}}\! - \! m_{\text{e}}\right)
&V_{22}	\left(m_{\text{s}}\! - \! m_{\mu }\right) 
&V_{23}\left(m_{\text{s}}\! - \! m_{\tau }\right)
\\
V_{31} \left(m_{\text{b}}\! - \! m_{\text{e}}\right)
&V_{32}
	 \left(m_{\text{b}}\! - \! m_{\mu }\right)
&V_{33} \left(m_{\text{b}}\! - \! m_{\tau }\right) \\
\end{pmatrix}\,,
\label{YukawaFromSymmetric}
\end{align}
where $V_{ij}$ denotes the elements of the $V_d$ mixing matrix.
 Clearly, the requirements like \cref{muonOut} are in contradiction with the unitarity of $V_d$ and thus LFV is principally unavoidable. 
 In \cite{Faber:2018qon} it was found (by scanning over the considered parameter space) that the experimental bound BR$(K_L\rightarrow \mu e)< 4.7 \times 10^{-12}$ \cite{PDG2018} inevitably leads to $R_K\geq1$, at odds with measurements.

Consequently, this implies that the assumption
$V_d=U_d^*$ is inconsistent with the requirement of simultaneously
explaining $R_K$ and respecting the bound from the $K_L\to\mu e$ decay.
However, such a model assumption is only fully justified
at the scale where one still has the left-right symmetry which, however,
is broken  well above $v_\chi$ (see e.g.~\cite{Bertolini:2013vta} and refs. therein for explicit
constructions), and renormalization group effects will lead to a breaking of $V_d=U_d^*$ anyway.
We also note that the current model might not emerge
from $SO(10)$ but from another framework. 

In the general case of $V_d \neq U_d^*$ we have the freedom to choose 6 angles and 12 phases.
In order to suppress the muon number violating processes 
    we require the conditions~\eqref{muonOut} to be satisfied to a high precision (at least to the order of $m_e/v_\mathrm{ew
} \cos\beta$), especially due to the very stringent limits on $\mu \to e \gamma$, $\mu \to eee$ and $K_L\to \mu e$. The general form of $U_d$ and $V_d$ conforming  this requirement can be found in Appendix~\ref{app:sweet_spot}.

Two of the remaining three angles therein can be subsequently constrained by invoking the stringent upper limits on the lepton flavour violating $\tau$ decays, along with the desire to maintain non-negligible $\Delta R_{K^{(*)}}$. In the part of the parameter space with the best potential to fulfill these requirements the  Yukawa matrix of Eq.~\eqref{YdeSU4} takes the form
\begin{align}
\hat  Y_4^{de} &\simeq  \frac{\sqrt{{3}/{2}}}{v_\mathrm{ew} \cos\beta} 
\begin{pmatrix}
	0& 0 & m_\tau  e^{i\delta_4}\sin \phi\\
    m_\mathrm{s} e^{i\delta_1}/\sqrt{2} & 0& m_\tau  e^{i\delta_5}\cos\phi\\
    m_b  e^{i\delta_2}/\sqrt{2} & 0& -m_b  e^{i\delta_3}/ \sqrt{2}
\end{pmatrix}\,,
\label{Y4tuned}
\end{align}
as detailed in Appendix~\ref{app:sweet_spot}. It is parametrized by a single angle $\phi\in\langle 0,\pi\rangle$ and five phases $\delta_i \in \langle 0, 2\pi\rangle$ which, in turn, define what we call the "sweet spot" region.

In this part of the parameter space the experimentally preferred values of $R_{K}$ and $R_{K^*}$ call for
\begin{align}\label{eq:masscos}
m_{R_2} \cos\beta \simeq 20\, \mathrm{GeV},
\end{align}
regardless of the choice of $\phi$ and $\delta_i$'s. Hence, one needs $\cos\beta \ll 1$ in order to obey the bounds from direct leptoquark searches.
Since this requirement, together with \cref{Y4tuned}, selects a rather special part of the parameter
space, the question arises in which other
channels  such a setting can be tested. There are essentially
two broad classes of these, namely, the low energy observables and the
LHC signals. We will focus here on the low energy part first
and discuss the collider aspects in the next section.

\subsection{Predictions and smoking gun signals}

\begin{table}\begin{center}
\begin{tabular}{|c|c|}
\hline
\multicolumn{2}{|c|}{Numerical input values}     \\
\hline
\hline
\rule{0pt}{10pt}$Y_2$ & $ \text{diag}(10^{-8}, 10^{-7}, 10^{-5}) $ \\
\hline
\rule{0pt}{10pt}$Y_5$ & $\text{diag} (10^{-2}, 5 \cdot 10^{-2}, 10^{-1}) $ \\
\hline
\rule{0pt}{10pt}$v_\chi$ & $ 4 \cdot 10^6 $~GeV  \\
\hline
\rule{0pt}{10pt}$m_A,\ m_{R_2} $ & $ 2 \cdot 10^5$~GeV, \ 1500~GeV  \\
\hline
\rule{0pt}{10pt}$\cos \beta$ & 0.02  \\
\hline
\end{tabular}
\end{center}
\caption{Summary of the sample input values used in the numerical  analysis of Sects.~\ref{sec:lowenergy} and~\ref{sec:collider} (unless stated otherwise).
Note also that all other BSM scalars have masses of the order $\mathcal{O}(m_A)$.
}
\label{tab:parameters}
\end{table}

From the construction it is clear that there will
be no additional constraints from any muon number violating
decays such as $\mu\to e \gamma$. 
In fact, we can achieve any value of BR($\mu\to e \gamma$) between zero and the experimental bound by arranging small deviations from the extreme scenario \cref{muonOut} with essentially no impact on the findings below.
In contrast, at the same time we cannot avoid sizeable effects in the $\tau$ sector. 
\subsubsection{Tau decays}
\begin{figure}
\begin{center}
\includegraphics[scale=0.37]{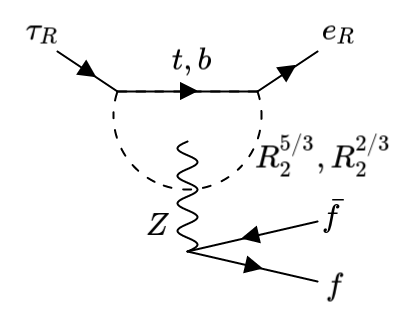}
\caption{The dominant interactions mediating the lepton flavour violating $\tau$ decays. }
\label{fig:penguin}
\end{center}
\end{figure}

The leptoquarks contribute to the $\tau \to e\gamma$ and $\tau \to e e e$ decays
at the loop level whereas to the final states involving mesons  
already at the tree level. Nevertheless, we find that, due to the differences in the magnitudes of relevant Yukawas, the $Z$-penguins (see \cref{fig:penguin}) induced by the third generation quarks dominate
over the tree-level contributions also for the $e\pi^+ \pi^-$ and $e K^+ K^-$ final
states.

A somewhat less important contribution arises from the photon penguin,
where in \cref{fig:penguin} the $Z$ gets replaced by $\gamma$. 
We have collected the relevant formulas  for the $\gamma$- and $Z$-contributions 
in \cref{sec:taudecay}. We find that the photon contribution comes
with an overall factor $1/m^2_{R_2}$ and that of the $Z$-penguin with a factor $m^2_Z$. We note for completeness that, nevertheless,  the structure
of the loop-functions is such that all these vanish in
the limit of $m_{R_2}\to \infty$. We have also found that the box contributions are subdominant.

For the evaluation of the predictions we 
have extended the \texttt{Flavor-Kit} package \cite{Porod:2014xia}
to include decays of the $\tau$ lepton into an electron
and two mesons.
For the calculation of BR$(\tau \to e \pi^+ \pi^-)$ we have adopted the formulas of refs.~\cite{Arganda:2008jj,Lami:2016vrs}. In the case of 
the $e \pi^\mp K^\pm$ final states
we have taken the form factor given in
\cite{Jamin:2008qg} but used updated values for the corresponding
meson masses.  For the calculation of the $B$-meson
observables we have employed
\texttt{flavio} \cite{Straub:2018kue} to which the data have
been transferred via the \texttt{wcxf}-interface \cite{Aebischer:2017ugx}.

\begin{figure}
\begin{center}
  \includegraphics[scale=0.66]{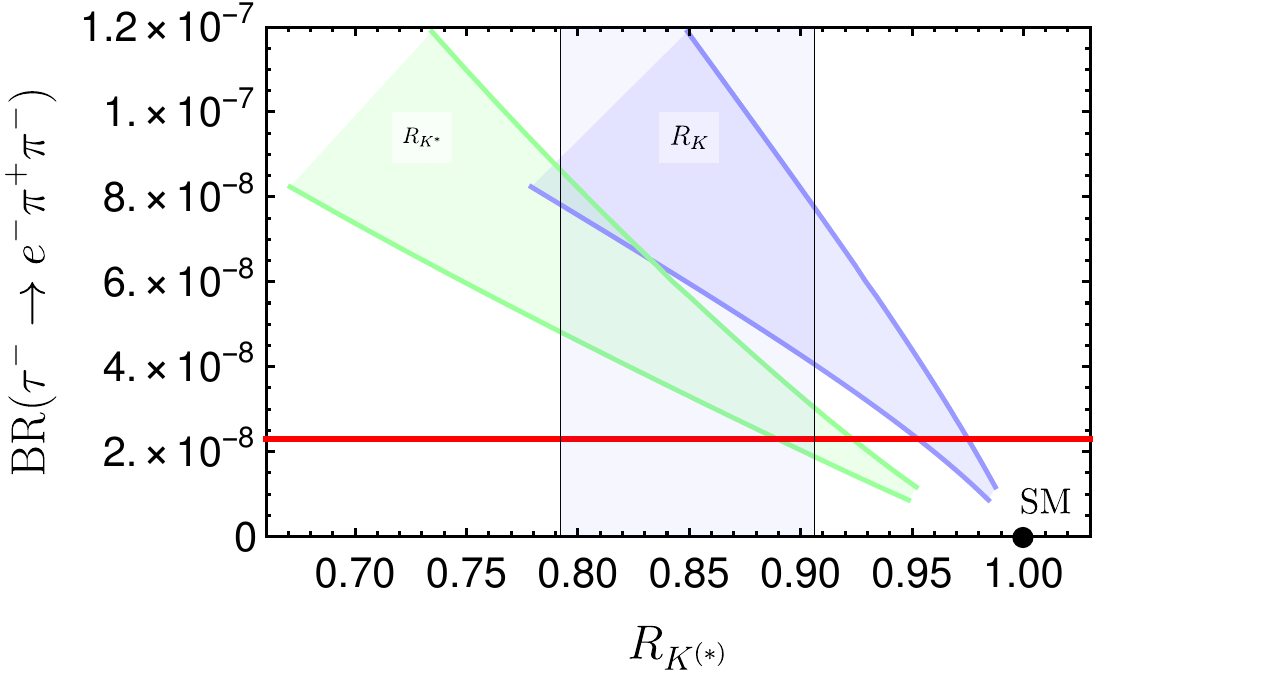}
\end{center}
    \caption{Correlations between BR($\tau \to e \pi^+ \pi^-)$ and $R_K$ as well as $R_{K^*}$. The input parameters have been chosen as in 
    \cref{tab:parameters} but $m_{R_2}$ has been varied between 0.9 and 2~TeV. Moreover, the quark
    mixing angles have been varied within the
    sweet spot region given by Eq.~\eqref{Y4tuned} yielding
    the bands shown.   
    The red horizontal line gives the current bound
    $\mathrm{BR}(\tau\to e \pi^+ \pi^-) \le 2.3 \cdot 10^{-8}$ whereas the vertical band 
    indicates the currently preferred  range of  ref.~\cite{Aaij:2019wad}: 
    $0.792 \le R_K \le 0.906$. 
    }
 \label{fig:tau_versus_RK}   
\end{figure}

%
The most stringent additional constraint stems
from  $\tau \to e \pi^+ \pi^-$.
In \cref{fig:tau_versus_RK} we show
BR$(\tau \to e \pi^+ \pi^-)$ versus $R_K$ where we have taken
the parameters as given in \cref{tab:parameters} except for
$m_{R_2}$, which we have varied from 900 GeV to 1.5 TeV, and scanned over all the sweet spot parameters of Eq.~\eqref{Y4tuned}.
Note that any digression from the sweet spot setting, i.e., diverting $\alpha$ and $\gamma$ of Appendix~\ref{app:sweet_spot} from their optimal values (\ref{eq:sweetspotalphagamma}), generally pushes the resulting BR($\tau \to e \pi^+ \pi^-)$  up.
One observes that there is a clear tension between the currently preferred
value for $R_K$ (cf.~\cref{eq:masscos}) and the bound on BR$(\tau \to e \pi^+ \pi^-)$
which excludes scenarios with $m_{R_2} \cos\beta \lesssim 30$~GeV. Hence, in what follows we adopt the limit case  $m_{R_2} \cos\beta = 30$~GeV (see Tab.~\ref{tab:parameters}) and calculate predictions for the other relevant decay rates of $\tau$-lepton, as summarized in Tab.~\ref{tab:tau_constraints}. 

The branching ratio for the final state containing $K^+ K^-$ is  smaller
by roughly a factor of~2 whereas those with
$\pi^\pm K^\mp$ are significantly
more suppressed due to the flavour-conserving quark current coupled to $Z$ in the relevant penguin.
We also find that the other flavour violating $\tau$ decays are
close to their experimental bound and within the sweet spot region vary only in a small range, see \cref{tab:tau_constraints}.
The dominance of the above
mentioned $Z$-penguin shows up also in the
predictions for BR($\tau \to e \mu^+ \mu^-$)
and BR($Z\to e^\pm \tau^\mp$) which vary
in the ranges $(1-1.3) \times 10^{-9}$ and
$(3.4-4.4)  \times 10^{-9}$, respectively.
We note that the predicted range
 BR($\tau\to e \gamma$) of a few times $10^{-9}$
provides a test of the current scenario at 
Belle II which aims to improve the limit on this channel to $3.3 \times 10^{-10}$ \cite{Kou:2018nap}. 
The model thus predicts that some lepton flavour
violating $\tau$-decays should be discovered soon.

We note for completeness that in the allowed parameter space the flavour violating $\tau$ decays into muons are strongly suppressed and, thus,
the observation of $\tau\to 3 \mu$   would rule out this scenario. 

\begin{table}[t]
 \centering
  \begin{tabular}{|c|c|c||c|c|c|}
    \hline
    $X$ & bound & range &  $X$ & bound & range \\ \hline
    $e\gamma$ &  $3.3 \cdot 10^{-8}$ & 
       $3.1 \cdot 10^{-9}$ - $3.8 \cdot 10^{-9}$ &
    $ee^+ e^-$ &  $2.7 \cdot 10^{-8}$ & 
      $1.2 \cdot 10^{-9}$- $1.6 \cdot 10^{-9}$ \\
    $e\pi^0$ & $8 \cdot 10^{-8}$ &
       $1.4 \cdot 10^{-9}$ - $4 \cdot 10^{-9}$ &
    $e\pi^+ \pi^-$ & $2.3 \cdot 10^{-8}$ &
      $1.9 \cdot 10^{-8}$ - $2.8 \cdot 10^{-8}$ \\
    $e K_S$ &  $2.6 \cdot 10^{-8}$ & 
      $7.7 \cdot 10^{-11}$ - $5.8 \cdot 10^{-11}$ & 
    $e K^+ K^-$ &  $3.4 \cdot 10^{-8}$ &
    $5.9 \cdot 10^{-9}$ - $8.5 \cdot 10^{-9}$ \\ 
%
    $e\phi$ &  $3.1 \cdot 10^{-8}$ &
      $1.2 \cdot 10^{-9}$ - $1.9 \cdot 10^{-9}$  &
        $e \pi^+ K^-$ &  $3.7 \cdot 10^{-8}$ & 
    $1.2 \cdot 10^{-20}$ -  $2.3 \cdot 10^{-11}$
    \\ \hline
    \end{tabular}
    \caption{Experimental bounds on various branching ratios BR($\tau \to X$)  \cite{PDG2018}
    and corresponding ranges in the sweet spot
    region, see main text, for the parameters given in \cref{tab:parameters}.}
    \label{tab:tau_constraints}
\end{table}

\subsubsection{Meson decays and oscillations}
We have also checked that the prediction for meson decays 
like $b\to s \gamma$, $B\to K \tau e$ , $B\to K \tau^+ \tau^-$ or $B_s\to \tau^+ \tau^-$
are fully consistent with the current experimental data.
In the context of leptoquarks a potentially constraining observable is
the ratio BR($K^+ \to e^+ \nu$)/BR($K^+ \to \mu^+ \nu$). However, due to the
required  smallness of $Y_2$, all leptoquark effects on observables with neutrinos in the final state are suppressed and, thus, also this is consistent with data. 

Staying in this part of the parameter space we have also checked whether the low energy
data can constrain the masses of the other components of $\Phi$. Our construction
implies that the scalar gluons, both the charged and the neutral one, 
have flavour mixing couplings to quarks. This means in particular that the neutral
one contributes at the tree level to $K^0$--$\bar{K}^0$ and $B_q$--$\bar{B}_q$ ($q=d,s$) mixing.
We find that within the experimental and theoretical uncertainties $B_s$--$\bar{B}_s$
requires $m_{G^0} \gsim 10$~TeV whereas in the case of the $K^0$--$\bar{K}^0$ mixing the bound is $m_{G^0} \gsim 15$~TeV.
It might be surprising that the $K^0$--$\bar{K}^0$ mixing limit is only slightly more stringent
than the $B$-meson one; this is a consequence of the specific shape of the parameter space  considered here. We have also checked that loop-induced contributions to the
$\Delta F=2$ transitions do not
provide additional constraints on the allowed parameter space.
We note, for completeness, that in other
parts of the parameter space this bound increases up
to $m_{G^0} \gsim 120$~TeV. 

\section{Collider phenomenology}
\label{sec:collider}
\subsection{Collider phenomenology in the presence of flavour anomalies}
\label{subs:colliderAnomalies}

In the previous section we have found a restricted region of the parameter space 
where a significant effect in $R_{K^{(*)}}$ can be accommodated while staying consistent with the constraints from other flavour observables such
as $\mu \to e \gamma$ and $K_L \to e \mu$.  Here we shall discuss interesting collider
signatures emerging in this part of the parameter space. Note that
\cref{massSumRule} allows for the situations where, apart from $R_2$, also the scalar gluons  $G$, or even the whole scalar sector arising from $\Phi$, can be light enough
to be tested either at the LHC or a prospective 100 TeV $pp$-collider.
 
Remarkably enough, in the slice of the parameter space  under consideration the leptoquarks
have rather special properties. In particular, the pattern of their Yukawa couplings \eqref{Y4tuned} is reflected in
their decays. For the charge $2/3$ particle one finds, regardless of which point in the sweet spot region is chosen,
\begin{align} 
\text{BR}( R_2^{+2/3} \!\to\! e^+ b) \simeq
\text{BR}( R_2^{+2/3} \!\to\! \tau^+ b) \simeq
\frac{m_b^2}{2m_\tau^2}\left(
\text{BR}( R_2^{+2/3} \!\to\! \tau^+ d) +  
\text{BR}( R_2^{+2/3} \!\to\! \tau^+ s) \right)\,,
\end{align}
where $\frac{m_b^2}{2m_\tau^2}\simeq 1.17$ is calculated at the  scale  $m_{R_2}$.
All other decay channels into charged leptons are negligible. Numerically, the BR's above amount to roughly 35~\% for the $e+j_b$ and $\tau+j_b$ final states and some 30~\% for the $\tau+j_\mathrm{light}$ in the case of $Y_2\simeq 0$, and scale down appropriately if $R_2^{+2/3}$ might decay into other channels like, e.g., $R_2^{+2/3}\to \bar\nu t$ due to possible non-zero entries in $Y_2$.

Due to the hierarchical structure of the CKM matrix, a similar pattern appears for the charge $5/3$ particle where, 
in the case of $Y_2\simeq 0$, the non-negligible decay channels satisfy
\begin{align} 
\text{BR}(R_2^{+5/3} \!\to\!  t e^+ ) \simeq
\text{BR}(R_2^{+5/3} \!\to\! t \tau^+) \simeq
\frac{m_b^2}{2m_\tau^2}\left(\text{BR}(R_2^{+5/3} \!\to\! u \tau^+) +  
\text{BR}(R_2^{+5/3} \!\to\! c \tau^+) \right)
\,.
\label{eq:BRfivethird}
\end{align}

These particles are
searched for by the ATLAS \cite{Aaboud:2016qeg} and 
CMS \cite{Sirunyan:2018vhk} experiments. Assuming
branching ratios of 100 \% into a specific channel such as $\tau b$,  bounds 
 up to 1.1~TeV have
been set if the leptoquarks are pair-produced. 
Since, however, various combinations
of different decay channels involving different
generations of fermions are allowed in the current scenario,  the actual
bounds are somewhat weaker. We have implemented
both analyses in the \texttt{CheckMate} framework \cite{Drees:2013wra,Dercks:2016npn} and found that
collider searches constrain the $R_2$ mass only
to about 890 GeV. This
clearly shows that the setting discussed in
\cref{sec:lowenergy} is fully consistent with bounds
from direct searches. 
 The decays of the
$R_2^{+5/3}$ to $t$-quarks give rise to missing energy
if the $W$ stemming from the $t$ decays leptonically. Therefore, we
checked in addition whether any of the SUSY searches implemented in
\texttt{CheckMate} can constrain our scenario. We find that
although some of the analyses do indeed show some
sensitivity to
the corresponding final states, they do not
exclude the current scenario. Potentially the high-luminosity
phase of LHC may further constrain it but this requires
a detailed study which is beyond the scope of this
paper.

We now turn to the next component of $\Phi$ which
can be potentially light, namely, the doublet of charged and neutral scalar gluons.
In what follows we will neglect the splitting of $G^0$ into its scalar and pseudoscalar component since it
is at most of $O(\mathrm{GeV})$.
The scalar gluon interactions arising from $Y_4$ generally read
\begin{align}
\label{eq:Glagrangian}
    \mathcal{L}_G & =
    \left[ 
    G^0 \overline{\hat d_L} 
+
    G^+ \overline{\hat u_L} 
    V_\mathrm{CKM} 
    \right]
       \hat Y^{dd}_4
    \hat d_R 
+\mathrm{h.c.}\,,
\end{align}
where the relevant Yukawa matrix satisfies 
\begin{align}
\hat Y^{dd}_4 =  \hat Y_4^{de} U_d^\dagger=    \frac{\sqrt{3/2}}{v_\mathrm{ew} \cos\beta} 
    \left(    \hat M_d -  V_d \hat M_e U_d^\dagger     \right)\,.
\label{YddSU4}
\end{align}
\begin{figure}[t]
    \centering
    \includegraphics[scale=0.4]{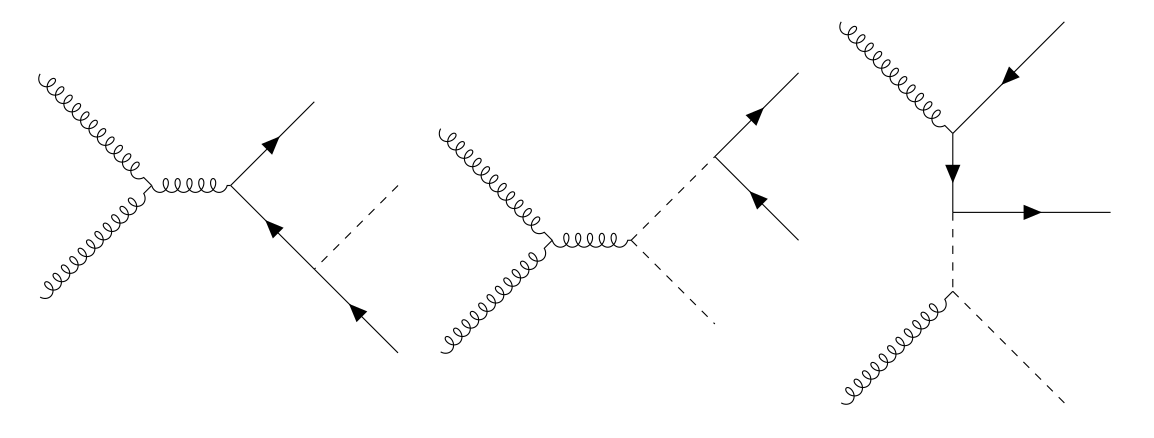}
    \caption{Exemplary Feynman graphs for the dominant production
    cross sections $pp\to G^0 q \bar{q}$ ($q=b,t$) and 
    $pp\to G^+ b \bar{t}$ at the hadron colliders.}
    \label{fig:FD}
\end{figure}%
Note that the interactions of the scalar gluons with right-handed up-type quarks originate from $Y_2$ which, as mentioned earlier,  is suppressed in our model. 
For this reason our findings  differ significantly from the ones of
refs.~\cite{Popov:2005wz,Frolov:2016gvu,Martynov:2017xws}.
Due to the $m_b$ enhancement in \cref{YddSU4}, the neutral scalar gluons are generally predicted to prefer decays to the $b$-quarks. 
In the sweet spot region discussed so far we obtain 
\begin{align}
\hat Y^{dd}_4 \simeq    \frac{\sqrt{3/2}}{v_\mathrm{ew} \cos\beta} 
\begin{pmatrix}
0 &m_\tau \sin\phi/\sqrt{2} & -m_\tau \sin\phi/\sqrt{2}\\
0& m_\tau \cos\phi/\sqrt{2} &-m_\tau \cos\phi/\sqrt{2} \\
-m_\mu  & 0 & m_b
\end{pmatrix} \,\,.
\label{Ydd}
\end{align}
Here we have neglected all phases as their
impact on the two-body decays is negligible.
One finds the following ranges for the various
branching ratios:
\begin{align}
\mathrm{BR}(G^0 \to b\bar{b}) &\simeq 0.7-0.75\,, \\
\mathrm{BR}(G^0 \to b\bar{d} +  d\bar{b}) +
\mathrm{BR}(G^0 \to b\bar{s} +  s\bar{b})&\simeq 0.15 \,,\\
\mathrm{BR}(G^+ \to t \bar{b}) &\simeq 0.65-0.73 \,.
\end{align}
The neutral states have also 
loop induced couplings to the gluons \cite{Gresham:2007ri}.
Denoting the scalar (pseudoscalar) component
of $G^0$ by $\sigma^0$ ($\phi^0$)  we find BR$(\sigma^0\to gg) \simeq 0.05$ and 
BR$(\phi^0\to gg) \simeq 0.01$. It has been noted already in ref.~\cite{Gresham:2007ri} that the scalar contributions in 
the loop induced couplings are negligible even for $\lambda_i=1$
and, thus, the parametric uncertainties due to the unknown
$\lambda_i$ are tiny. The remaining decays are into two quarks of the first two generations.
We found in the previous 
section that the mass of the scalar gluon
should be above $\sim 15$~TeV due to the constraints on the
$K^0$--$\bar{K}^0$ mixing. This is clearly too heavy
for the LHC and, thus, one needs a 100 TeV
$pp$-collider \cite{Arkani-Hamed:2015vfh,Contino:2016spe}
to look for these states. 

\begin{figure}[t]
\begin{center}
\includegraphics{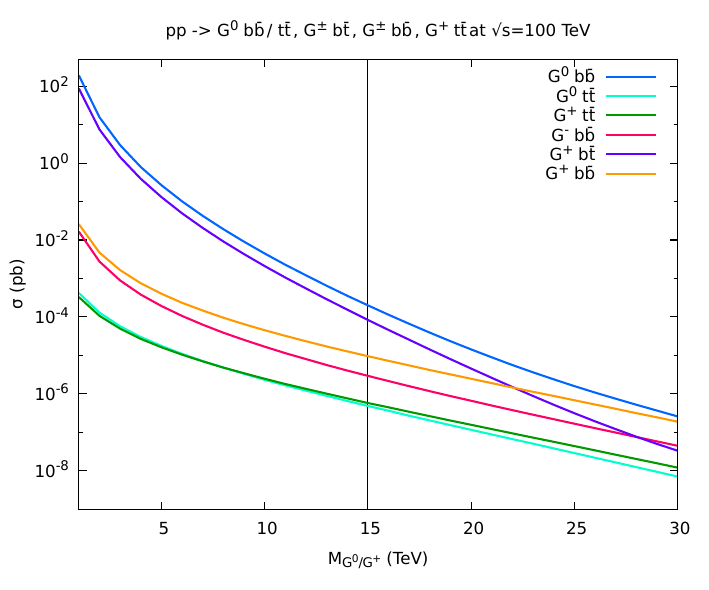}
\end{center}
\caption{\label{fig:SigmaG0G+_100TeV} Various production cross sections
at a prospective $pp$-collider with $\sqrt{s}=100$~TeV as a function
of the corresponding mass. In addition also the channel $G^- \bar{b} t$ 
exists and neglecting
the electroweak contributions one finds $\sigma(G^- \bar{b} t)=\sigma(G^+ \bar{t} b)$. Here we have used the parameters given in \cref{tab:parameters} except for
the masses of the scalar gluons. The vertical line indicates the bound on 
$m_{G^0}\simeq m_{G^+}$ obtained from
meson mixing. 
}
\end{figure}

In \cref{fig:FD}
we present some of the dominant Feynman diagrams
for the processes $pp\to G^0 q \bar{q}$ ($q=b,t$) 
and  $pp\to G^+ b \bar{t}$,
including also the contributions from the production of a scalar gluon pair with the subsequent decay of one of the scalar gluons into $q\,{\bar q} ^({} '{} ^)$.  
The corresponding cross sections for a 100 TeV collider
are shown in \cref{fig:SigmaG0G+_100TeV} where we have included all tree-level QCD contributions as well as all couplings of scalar gluons to quarks. The relevant Yukawa coupling $Y_4$
is choosen 
in the sweet spot region with $\phi = 0$.
For large scalar masses the production cross sections get a significant contribution also
from the quark initial states or are even dominated by those.
For instance,
$\sigma(pp \to G^+ b\bar{b})$ varies by about 20 per-cent within the sweet spot region because of its 
dependence on $Y_4$.
Note that the cross sections
shown here are calculated at the tree level and we expect sizeable QCD corrections. Combining the cross sections with the branching ratios above, we
have found that the dominant signals will be in the 4 $b$-jet and 2$t$+2$b$-jets channels
which are experimentally challenging. 

\subsection{Scalar gluons at colliders without flavour anomalies}
\begin{figure}[t]
\includegraphics[scale=0.67]{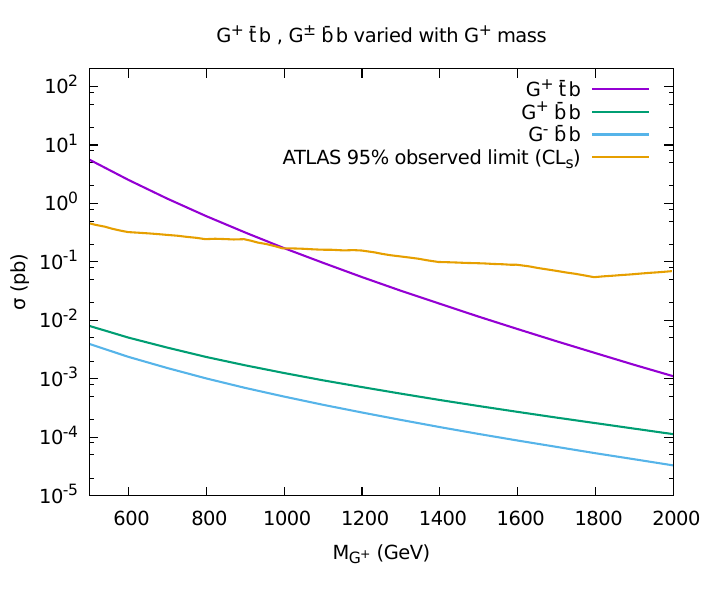}
\includegraphics[scale=0.67]{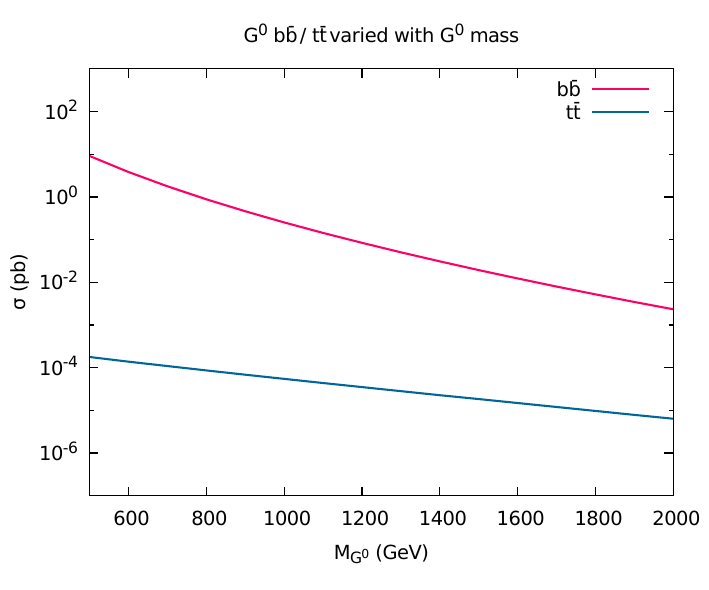}
\caption{Production cross sections
at the LHC with $\sqrt{s}=13$~TeV. On the left side the cross
sections $\sigma(pp \to G^+ \bar{t} b)$ (purple line), $\sigma(pp \to G^+ \bar{b} b)$ (green line) 
and $\sigma(pp \to G^- \bar{b} b)$ (blue line) are shown as a function of $m_{G^+}$.
The yellow line gives the current bound on the $G^+ \bar{t} b$ final state
obtained by the ATLAS experiment \cite{Aaboud:2018cwk}.
On the right side the cross sections 
$\sigma(pp \to G^0 \bar{b} b)$ and 
$\sigma(pp \to G^0 \bar{t} t)$ are shown as a function of $m_{G^0}$. }\label{fig:SigmaG0G+}
\end{figure}
Since the measurements of the $B$-anomalies still admit the case of being pure statistical fluctuations, in what follows we focus for completeness on the situation when both leptoquarks are too heavy to contribute significantly to the low energy observables and when the lightest BSM fields are the scalar gluons. 

These particles are interesting on their own, and, thus
we study here the limit case where all flavour
violating couplings of the neutral scalar gluons
are absent. This can be achieved, e.g.,
by setting $V_d$ and $U_d$ 
to the unit matrix which leads to
\begin{align}
\label{eq:Y4simplified}
\hat Y^{dd}_4 =  \sqrt{\frac{3}{2}}  \frac{1}{v_\mathrm{ew} \cos\beta} 
    \left(    \hat M_d -  \hat M_e   \right)\,.
\end{align}
Assuming that the elements of 
$Y_2$ are smaller at least by an order of magnitude compared to those in $Y_4$, the scalar gluons
can have masses within the reach of the LHC.
Consequently,  \cref{eq:Y4simplified}  together with \cref{eq:Glagrangian}
imply that BR$(G^+\to t \bar{b})$ is close to one and that 
the neutral states decay dominantly into $b\bar{b}$.  

The latter can also decay into two gluons.
However, in this setting the neutral scalar gluons
have suppressed couplings to the top-quark
compared to the situations  discussed for example
in refs.~\cite{Smirnov:1995jq,Gresham:2007ri,Hayreter:2018ybt}, which is due to the smallness of the $Y_2$ entries in the current scenario. Consequently, also the loop-induced
$G^0 g g$ coupling is significantly smaller compared to models
where $Y_2$ induces large couplings to the $t$-quark.
Firstly, this implies that the decays
into two gluons have a branching ratio of at most 5 per-cent. 
Secondly, this also implies that the bounds 
from processes like
\begin{align}
pp \to G^0 + X \to 2 j + X
\end{align}
(with $j$ being either a quark or a gluon jet) obtained by the CMS experiment \cite{Sirunyan:2018pas,Sirunyan:2018xlo} do not
constrain our model even when taking
QCD corrections via a K-factor of 1.7 
\cite{GoncalvesNetto:2012nt} into account.
Instead, we have found that the strongest
constraints come from the ATLAS
search for the $H^+ \bar{t} b$ production
\cite{Aaboud:2018cwk}. We can see from 
\cref{fig:SigmaG0G+} that this excludes
scenarios with $m_{G^+} \simeq 1$~TeV. 
This is actually a conservative bound in the sense that we assume
here BR($G^+\to t \bar{b})=1$ which maximizes the power of the
experimental analysis. 
We want to stress that we have also
included here the pair production $pp\to G^+ G^-$
combined with the subsequent decay 
$G^- \to \bar{t} b$. Due to the steep decrease of the cross sections with
the mass this plot indicates that the reach of the LHC will not be above
1.5 TeV. We therefore show in \cref{fig:SigmaG0G+_100TeV} 
various cross sections at the 100 TeV collider starting from masses in the
TeV range which clearly shows that the cross sections in the low mass range is 
so large that these particles should be found within the first data sets.

\section{Conclusions}
\label{sec:conclusions}

In this paper we have studied a model based on the extended $SU(4)_C\times SU(2)_L \times U(1)_R$ gauge symmetry which is arguably the most minimal UV-complete gauge framework including vector and scalar leptoquark fields.  It has been shown recently \cite{Faber:2018qon} that, among other features,  this  setup has the potential to
accommodate significant effects in semileptonic $B$-decays. It is well known that, in this context, the strongest
constraints stem from the non-observation of
$K_L \to e \mu$ and $\mu \to e \gamma$. In order for these to be satisfied 
along with $R_{K^{(*)}}$ well below 1,
a~rather specific flavour pattern of the scalar leptoquark interactions with matter is required; for instance, all couplings of the supposedly light $R_2$ leptoquark to muons need to be strongly suppressed along with its couplings to the right-handed up-type quarks and left-handed charged leptons.  
We have shown that there exists a narrow region in the parameter space where a highly consistent picture can be achieved. This, in turn, leads to a very predictive scenario in which several other  interesting phenomenological conclusions can be drawn.

First, there are sharp predictions for the branching ratios of $\tau\to e \pi^+ \pi^-$ and  
$\tau\to e K^+ K^-$ which are marginally compatible with the
current experimental bounds; especially the first channel provides a strong constraint on the allowed
parameter space. 
Moreover, also
$\tau\to e \gamma$ and $\tau \to 3 e$  turn out to be 
close to their current experimental limits and, thus, their signals should be observable in the next round of experiments such as Belle II. Thus, if $R_K$ stays on its current value, non-observation of the LFV $\tau$-decays will rule out the model as a whole. 

Second, the charge-$2/3$ and $5/3$ scalar leptoquarks, whose masses should not be much above 1 TeV in order to address the $B$-anomalies, turn out to have
rather specific decay properties which can be tested either at LHC or
at a future 100 TeV $pp$ collider. In particular, we find that 
$\text{BR}( R_2^{+2/3} \to e \bar{b}) \simeq
\text{BR}( R_2^{+2/3} \to \tau \bar{b}) \simeq
\text{BR}( R_2^{+2/3} \to \tau \bar{d}) +  
\text{BR}( R_2^{+2/3} \to \tau \bar{s})$ and
$\text{BR}(R_2^{+5/3} \!\to\!  t e^+ ) \simeq
\text{BR}(R_2^{+5/3} \!\to\! t \tau^+) \simeq
\text{BR}(R_2^{+5/3} \!\to\! u \tau^+) +  
\text{BR}(R_2^{+5/3} \!\to\! c \tau^+) $. As such, a clear indication should be expected in the high-luminosity LHC run if the mass of $R_2$ was in the indicated 1.5 TeV ballpark.

Third, there is enough room in the allowed parameter space  for relatively light scalar gluons  (with electric charges 0 and 1) whose masses are constrained from meson mixing to be above some 15~TeV. Again, the branching ratios of their decays (including those into  flavour violating channels) are fixed within
narrow ranges which would facilitate their searches at  future colliders.

Remarkably enough, the phenomenology of such relatively light scalar gluons in the model under consideration is interesting even if the $B$-anomalies eventually disappear. It turns out that in such a case the stringent limits from the meson mixing can be alleviated and the bounds on their masses can be lowered into the LHC domain.
In this scenario the most stringent limit stems from
the process $pp\to G^+ \bar{t} b$ where we get
a bound $m_{G} \simeq 1$ TeV recasting  an ATLAS
search for $H^+$. The usual bounds on $G^0$ do not
apply in this model. In that situation the branching ratios into the third generation quarks, namely, BR($G^+\to t \bar{b}$) and BR($G^0\to b \bar{b}$), 
amount to almost 100 \%.  

\section*{Acknowledgments}
We thank Jorge Portol\'ez and Maria Jose Herrero Solans 
for useful discussion on $\tau$ decays into an $e^-$ and two mesons  and Vojt\v{e}ch Pleskot for comments on the LHC leptoqurk production.  
F.S.\ is supported by the ERC Recognition Award ERC-RA-0008 of the Helmholtz 
Association. T.F., Y.L.\ and W.P.\ have been supported by  the DFG, project 
nr.\ PO-1337/7-1.
M.H.\ and M.M.\ acknowledge the support from the Grant agency of the Czech Republic, project no.~20-17490S and from the Charles University Research Center UNCE/SCI/013. M.H. has also been supported by the Grant Agency of Charles University (GAUK) project no. 12481/2019. 
 H.K.\ has been supported by the
grant no.\ PR-10614 within the ToppForsk-UiS program of the University of
Stavanger and the University Fund. Finally, we acknowledge the comments of the referee which helped to discover an error in the original numerical code.  

\appendix

\section{The "sweet spot" region}
\label{app:sweet_spot}
We approximate Eq.~\eqref{YdeSU4} by neglecting $m_e$ and $m_d$, and further by neglecting 
the second generation masses when compared to the third ones. The resulting structure, omitting the $\sqrt{3/2}\, ({v_\mathrm{ew} \cos\beta})^{-1} $ prefactor, reads
\begin{align}
\hat Y_4 ^{de}
\propto
\begin{pmatrix}
 0
& - V_{12} m_{\mu }
& - V_{13}   m_{\tau }
\\
U_{21} m_{s}
&U_{22}	m_{s}\! - V_{22} m_{\mu }
&-V_{23} m_{\tau }
\\
U_{31} m_{b}
&V_{32} m_{b}
&U_{33} m_{b} - V_{33} m_{\tau }
\end{pmatrix}.
\label{YukawaApprox}
\end{align}
As explained in the main text, we need to set the middle column, corresponding to the muon interactions, to zero. Within the current approximation scheme this is achieved if and only if the mixing matrices take the form 
\begin{align}
U_d &=
\begin{pmatrix}
 e^{i \delta _8} \cos\gamma \sin \alpha & e^{i (\delta _7+\delta _8-\delta _1) } \cos \alpha & e^{i (\delta _3+\delta _8-\delta _2) } \sin \alpha \sin\gamma \\
 -e^{i \delta _1} \cos \alpha \cos\gamma & e^{i \delta _7} \sin \alpha & -e^{i (\delta _1-\delta _2+\delta _3) } \cos \alpha \sin\gamma \\
 -e^{i \delta _2} \sin\gamma & 0 & e^{i \delta _3} \cos\gamma 
\end{pmatrix}\,, 
\label{Ud}
\\
V_d &=
\begin{pmatrix}
e^{i \delta _9} \cos \phi & 0 & -e^{i \delta _4} \sin \phi \\
 -e^{i (-\delta _4+\delta _5+\delta _9) } \cos \alpha  '  \sin \phi & e^{i \delta _7} \sin \alpha  '  & -e^{i \delta _5} \cos \phi \cos \alpha  '  \\
 e^{i (-\delta _4+\delta _6+\delta _9) } \sin \phi \sin \alpha  '  & e^{i (-\delta _5+\delta _6+\delta _7) } \cos \alpha  '  & e^{i \delta _6} \cos \phi \sin \alpha  '  \\
\end{pmatrix}\,,
\label{Vd}
\end{align}where $\alpha$ and $\alpha'$ are related via $m_s \sin\alpha  = m_b \sin \alpha' $. This yields
\begin{align}
\hat Y_4 ^{de} = \frac{\sqrt{3/2}}{v_\mathrm{ew} \cos\beta}
\begin{pmatrix}
 0 & 0 & e^{i \delta _4}  m_{\tau } \sin \phi \\
 -e^{i \delta _1}  m_{s}\cos \alpha \cos\gamma & 0 & e^{i \delta _5}  m_{\tau }\cos \phi \cos \alpha  '  \\
 -e^{i \delta _2} m_{b} \sin\gamma  & 0 & e^{i \delta _3}m_{b} \cos\gamma -e^{i \delta _6}  m_{\tau } \cos \phi \sin \alpha' 
\end{pmatrix}
 \label{Yde:sweetSpot}
\end{align}Note that only small $O(m_\mu/m_b)$ deviations from the displayed form of $U_d$ and $V_d$ are necessary in order to fulfill the condition \eqref{muonOut} exactly. We always use those exact forms in the numerical calculations (\SPheno) but, for the sake of clarity, stick within  the approximation in \eqref{Ud} and \eqref{Vd} in all the equations in the text.

Another important restriction on the parameter space emerges from  $\tau\to e \pi^+\pi^-$ in which the tree-level $R_2^{+5/3}$ leptoquark contribution is suppressed with respect to the top--$R_2^{+5/3}$ induced $Z$-penguin. The latter is driven by the product of the 31 and 33 elements of $(V_\mathrm{CKM}.\hat Y_4^{de})$ which, due to $V_{tb}\approx 1$, essentially coincide with the 31 and 33 elements of~(\ref{Yde:sweetSpot}). Hence, the ratio of the effective couplings governing $\Delta R_{K^{(*)}}$ and $\tau\to e \pi^+\pi^-$ is roughly proportional to
\begin{equation}
\label{eq:ratio}
\frac{C_{\Delta R_{K^{(*)}}}}{C_{\tau\to e \pi^+\pi^-}}\propto \frac{y_{be} y_{se}}{y_{be} y_{b\tau }}= 
\frac{-e^{i \delta _1}  m_{\text{s}}\cos \alpha \cos\gamma}{e^{i \delta _3}m_{\text{b}} \cos\gamma -e^{i \delta _6} \frac{m_s  m_{\tau }}{m_b} \cos \phi \sin \alpha}\,.   
\end{equation}
As long as $\cos \gamma$ is non-negligible, the first term in the denominator of~(\ref{eq:ratio}) dominates and the RHS therein depends only on $\cos\alpha$ and an irrelevant overall phase. Hence, in order to maximize the effect in $R_{K^{(*)}}$ one should keep $\alpha\approx 0$.
Note that the case of $\cos \gamma\sim 0$ is pathological as in this situation any sizeable effect in $R_{K^{(*)}}$ relies on further enhancing $(m_R \cos\beta)^{-1}$ which either renders the 31 coupling of~(\ref{Yde:sweetSpot}) non-perturbative (for tiny $\cos\beta$) or requires very low $m_{R_2}$, at odds with direct searches.

To conclude, we shall fix
\begin{equation}
\label{eq:sweetspotalphagamma}
\alpha=0\;\;,\;\gamma=\pi/4
\end{equation}
so that to maximize the effects in $R_{K^{(*)}}$.  The remaining  parameters are left free and span what we call the ``sweet spot" region.

\section[Dominant contributions to 
the decay tau -> e pi+ pi-]{Dominant contributions to 
the decay $\tau\to e \pi^+ \pi^-$}
\label{sec:taudecay}
We collect here the formulae for the photon
and the $Z$ penguins including leptoquarks for the
decay $\tau \to e \pi^+ \pi^-$. The corresponding
matrix elements are given for the quark currents
which then need to be hadronized according to the
procedure presented in \cite{Arganda:2008jj,Jamin:2008qg}.

The matrix elements for the photon contribution reads
\begin{align}
T_{\gamma,q} = \bar{u}_e(p_1)\left[k^2 \gamma_{\mu}  A_1  + i m_{\tau} \sigma_{\mu \nu} k^{\nu}  A_2 \right]P_R u_\tau(p)
\, \frac{e^2}{k^2} e_q \bar{u}_q(p_2) \gamma^ {\mu} v_q(p_3) \,,
\end{align}with
\begin{align}
A_1 &= \frac{N_c}{576 \pi^2} \sum_q  y_{qe}^* y_{q\tau}
\frac{1}{m^2_{R_2}} \left( f(x_q) + e_q g(x_q) \right) \,,
\\
A_2 &= \frac{N_c}{32 \pi^2} \sum_q y_{qe}^* y_{q\tau}
\frac{1}{m^2_{R_2}} \left( \tilde f(x_q) + e_q \tilde g(x_q) \right) \,,
\end{align}
where $k$ is the photon 4-momentum, 
$e_q$ is the charge fraction of the corresponding quark, 
$N_c=3$, $y_{ql}$'s are defined in \cref{Yde}
and we have neglected terms proportional to $m_e/m_\tau$.
The sum runs over all quarks, with $x_q=(m_q/m_{R_2})^2$. 
Note that in the case of $d$-type ($u$-type) quarks, 
$m_{R_2}$ denotes the mass of the charge $2/3$ ($5/3$) leptoquark. 

The loop-functions above take the form
\begin{align}
f(x) &= \frac{2 - 9 x +18 x^2-11 x^3+6 x^3 \log x}{\left(1-x\right)^4}\,,
\\
g(x) &= \frac{6 \left(3-3 x+ \left(2 +x\right) \log x \right)}{(1-x)^2}\,,
\\
\tilde f(x) &= \frac{1- 6 x+3 x^2 + 2 x^3-6 x^2 \log x}{6 \left(1-x\right)^4}\,,
\\
\tilde g(x) &= \frac{x^2-1 -2 x \log x}{2 \left(x-1\right)^3}\,.
\end{align}We have cross-checked that we can reproduce the formulas given
in \cite{Arganda:2005ji,Crivellin:2018qmi}. 

The
matrix elements for the $Z$-boson contribution read
\begin{align}
T_{Z,q} &= \frac{1}{m^2_Z}
\bar{u}_e(p_1) \gamma^\mu F_R P_R u_\tau(p)
\;  \bar{u}_q(p_2) \gamma_\mu (a^q_L P_L + a^q_R P_R) v_q(p_3)\,,
\end{align}where the loop-induced flavour violating $Z$-vertex is
\begin{align}\label{lifvZv}
F_R &= -\frac{N_c}{16 \pi^2} \sum_q y_{qe}^* y_{q\tau} 
\left(2 a^q_R C_{1F}(x_q) - a^q_L C_{2F}(x_q)
-2 a^{LQ} C_B(x_q) + a^l_R B_1(x_q)  \right)\,,
\end{align}
with  $a^q_{L,R}$, $a^l_R$ and
$a^{LQ}$ denoting the couplings of the $Z$-boson to
quarks, leptons and leptoquarks, $a^i=-\frac{g}{\cos\theta_W}  \left(T^3_i - e_i \sin^2\theta_W\right)$, respectively.

The loop functions in~(\ref{lifvZv}) read
\begin{align}
C_{1F}(x) &=\frac{1-x+\left(2-x\right) x \log x}{4 (1-x)^2}
 -\frac{1}{4} \log \left(\frac{m_{R_2}^2}{Q^2}\right)\,,   \\
C_{2F}(x) &= \frac{x-x^2+ x\log x}{(1-x)^2}\,, \\
C_{B}(x) &= 
\frac{x^2-x-x^2 \log x}{4   (1-x)^2}
   -\frac{1}{4} \log \left(\frac{m_{R_2}^2}{Q^2}\right)\,,\\
B_1(x) &= \frac{1}{4} \left(\frac{1-x^2+2 x^2 \log x}{(1-x)^2}+2 \log \left(\frac{m_{R_2}^2}{Q^2}\right)-2\right)
   \, .
\end{align}
We have checked that we can also reproduce the corresponding formulas
given in \cite{Arganda:2005ji}, apart from $B_1$ which
contains a trivial typo which can be easily seen
by noting that $B_1$ has to be dimensionless.
At first glance it might be surprising to
see an explicit dependence on the renormalisation
scale $Q$. However, within one quark species one
finds immediately that the terms proportional to
$\log(m^2_{R_2}/Q^2)$ yield
\begin{align}
\frac{y_{qe}^* y_{q\tau}}{2}
(- a^q_R + a^{LQ} + a^l_R) 
  \log\left(\frac{m^2_{R_2}}{Q^2}\right)  =
\frac{y_{qe}^* y_{q\tau}}{2} T^3_{LQ}  \log\left(\frac{m^2_{R_2}}{Q^2}\right)\,. 
\end{align}
Performing the sum over the $u$-type quarks,
the CKM-matrix drops implying this part of the $Z$-penguin
is indeed proportional to
\begin{align}\log\left(\frac{m^2_{R^{2/3}_2}}{m^2_{R^{5/3}_2}}\right)
 \sum_{q=d,s,b}^3 y_{qe}^*y_{q\tau}\,.
\end{align}
We note for completeness that the ratio of the two
leptoquark masses becomes 1 in the $SU(2)_L$ conserving limit in which this contribution vanishes as expected. 
Though not quite obvious from its form we note that $F_R$
also vanishes in the limit of infinite leptoquark masses.


\section{Implementation in \SARAH}
\label{app:SARAH}
\subsection{Changes in \SARAH}
In the context of this project, we have extended the functionality of \SARAH to work with unbroken subgroups in order to implement the 
Pati-Salam model. We summarize the main parts of the \SARAH model file and explain the new commands. For all details of the 
standard commands we refer to Refs.~\cite{Staub:2008uz,Staub:2015kfa}. The following changes in \SARAH have happened:
\begin{enumerate}
 \item The $SU(4)_C$ algebra was implemented to express the generators and structure constant of $SU(4)_C$ in terms of generators and structure constants of $SU(3)$ and Kronecker deltas. 
 \item The possibility to define unbroken subgroups of a bigger gauge group was added
 \item All necessary routines to write the matter and gauge fields, which are defined for the bigger group, in terms of the unbroken subgroup were developed
\end{enumerate}
We tried to keep the changes in \SARAH as generic as possible. I.e. the new functionality is not restricted to the considered model or to Pati-Salam groups. However, we have only tested the function thoroughly for the model discussed in this paper. Therefore, one should be careful when using it with other models. 

\subsection{The \SARAH model files}
\begin{enumerate}
\item The fundamental gauge groups ($SU(4)_{C}\times SU(2)_L \times U(1)_R$) are defined as usually via the array {\tt Gauge}:
\begin{MIN}
Gauge[[1]]={WR,U[1],right,gR,True};
Gauge[[2]]={WL,SU[2],left,gL,True};
Gauge[[3]]={PS,SU[4],pati,g3,True};
\end{MIN}
\item In order to define that $SU(4)_C$ get broken to an unbroken group $SU(3)$, the following three steps are necessary:
\begin{enumerate}
 \item The name of the group which shall be broken as well as the name of the unbroken subgroups are defined via {\tt UnbrokenSubgroups}
\begin{MIN}
UnbrokenSubgroups={pati->color};
\end{MIN}
Here, the first part of the rule must correspond to an entry in {\tt Gauge}. 
\item The features of the unbroken gauge groups in the new array {\tt AuxGauge} are defined. This is completely analogue to the definition of a group in {\tt Gauge}. 
\begin{MIN}
AuxGauge={{G,SU[3],color,g3,False}};
\end{MIN}
The third entry must be identical to the chosen name in {\tt UnbrokenSubgroups}. 
\item Names for the new gauge bosons must be introduced. The mapping between the fundamental gauge bosons ($V_1 \dots V_N$) to a set of new gauge bosons
$\{V^a,V^b,\dots V^x\}$ with dimensions $\{N_a,N_b,\dots N_b\}$ under the unbroken subgroup is done as
\begin{equation}
\left(\begin{array}{c} 
      V_1 \\
      V_2 \\
      V_3 \\
      {} \\
      {} \\
      \cdot \\
      \cdot \\
      \cdot \\
      {} \\
      {} \\
            V_{N-1} \\
      V_N
      \end{array}
\right) = 
\left(\begin{array}{c} 
      V^a_1 \\
      \dots \\
      V^a_{N_a} \\
      V^b_1 \\
      \dots \\
      V^b_{N_a} \\
      \cdot \\
      \cdot  \\
      \cdot \\
      V^x_1 \\
      \dots \\
      V^x_{N_a} \\
      \end{array}
\right) 
\end{equation}
This relation is defined in the model file using the new array {\tt RepGaugeBosons}. For each unbroken subgroup a list must be given which consists of pairs of the name of a gauge boson and its dimension.
\begin{MIN}
RepGaugeBosons = {{{VG,8}, {VX,3}, {VY,3}, {VS,1}}};
\end{MIN}
Note, the names for the gauge bosons must always start with {\tt V}. From this definition, also the mapping of the ghost is  derived. The names of the ghost fields are those of the vector boson with {\tt V}
replaced by {\tt g}. 
\end{enumerate}
\item After the definition of the gauge groups, the matter fields are defined. This is done for non-supersymmetric fields using the arrays {\tt FermionField} and {\tt ScalarField}. For fields which transform 
non-trivially under the broken gauge groups, the tensor notation is used. Thus, the fundamental representation is a vector of dimension $N$. If the unbroken subgroup has dimension $n$, the relation between the 
components of the fields are 
\begin{equation}
\left(\begin{array}{c} 
      \Phi_1 \\
      \Phi_2 \\
      \dots \\
      \Phi_{N-1} \\
      \Phi_N
      \end{array}
\right) = 
\left(\begin{array}{c} 
      \Psi_1 \\
      \dots \\
      \Psi_{n} \\
      \Psi' \\
      \dots  \\
      \Psi^{'\dots'} \\
      \end{array}
\right) 
\end{equation}
The number of fields with a prime is $N-n$. \\
For the adjoint representation, an $N\times N$ matrix is used. This matrix is then decomposed as
\begin{equation}
\left(\begin{array}{ccc} 
      \Phi_{11} & \dots & \Phi_{1N} \\
      \vdots & &  \vdots \\
      \Phi_{N1} & \dots & \Phi_{NN} 
      \end{array}
\right) = 
\left(\begin{array}{cccccc} 
      \Psi_{11} & \dots & \Psi_{1n} & \Psi'_{1} & \dots & \Psi^{'\dots'}_1  \\
      \vdots & & \vdots & \vdots & & \vdots \\
      \Psi_{n1} & \dots & \Psi_{nn} & \Psi'_{n} & \dots & \Psi^{'\dots'}_n\\
      \tilde \Psi'_{1} & \dots & \tilde \Psi'_{n} & \alpha' & \dots & \phi^{'\dots'}   \\
      \vdots & & \vdots & \vdots & & \vdots  \\
      \tilde \Psi^{'\dots'}_1& \dots & \tilde \Psi'_{n} & \omega'  & \dots & \omega^{'\dots'}    \\
      \end{array}
\right) 
\end{equation}
Here, $\Psi$ is in the adjoint  representation of the unbroken subgroup and all primed fields $\Psi'$ and $\tilde\Psi$ are vectors under the unbroken subgroup. The fields $\alpha$ to $\omega$ are singlets under the 
unbroken group. 
\begin{enumerate}
\item In the given model, the fermion fields are either singlets or transform in the (anti-) fundamental representation. This is defined via
\begin{MIN}
FermionFields[[1]] = 
  {FQL,3,{{uL[color,3],vL},{dL[color,3],eL}},0,2,4};
FermionFields[[2]] = 
  {FU,3,{uR[color,-3],vR},-1/2,1,-4};  
FermionFields[[3]] = 
  {FD,3,{dR[color,-3],eR},1/2, 1,-4};
FermionFields[[4]] = {Si,3,Sing,0,1,1};
\end{MIN}
Note, here the last three entries define the representation with respect to the gauge groups defined in {\tt Gauge}. The representation with respect to the unbroken subgroup are defined for each component field 
in squared brackets, i.e. {\tt uL[color,3]} means that the field {\tt uL} is a colour triplet. 
\item In the scalar sector, the adjoint representation is needed in addition. All scalars are defined via
\begin{MIN}
ScalarFields[[1]] =  
  {Chi,1,{Chiu[color,3],Chi0},1/2,1,4};
ScalarFields[[2]] = {H,1,{Hp,H0},1/2,2,1};
ScalarFields[[3]] = 
{Phi15,1,{
{{HGp15[color,colorb,8]+HSp15*dAB/Sqrt[12],HXp15[color,3]},{HYp15[colorb,3],-3*HSp15/Sqrt[12]}},
{{HG015[color,colorb,8]+HS015*dAB/Sqrt[12],HX015[color,3]},{HY015[colorb,3],-3*HS015/Sqrt[12]}}
},1/2, 2,15};
dAB=Delta[color,colorb]                             
\end{MIN}
Here, we have introduced the abbreviation {\tt dAB} only for better readability. Note, that for the tensor representation the name of the second colour index is extended by {\tt b} (i.e. {\tt colorb} to prevent 
any ambiguity). \\
There is one additional subtlety: in \SARAH and other codes like {\tt MadGraph}, {\tt CalcHep} or {\tt WHIZARD} the higher dimensional representations of {\it unbroken} gauge groups, i.e. the colour group, are not written as tensors but vectors. Therefore, it is necessary to re-write the neutral and charged octets. The necessary definitions are given in the list {\tt TensorRepToVector} which reads in our case:
\begin{MIN}
TensorRepToVector=
{
 {HG015,{color,HGV015,HG015[{p_,a_,b_}]:> 
   sum[color/.subGC[gNN[p]],1,8]
   Lam[color/.subGC[gNN[p]],a,b]
   HGV015[{p,color/.subGC[gNN[p]]/Sqrt[2]}], 
  {sum[col6,1,3]sum[col6b,1,3]Lam[col1,col6,col6b],
  {col1->col6,col1b->col6n}}}
 },
 {HGp15,{color,HGVp15,HGp15[{p_,a_,b_}] :> 
  sum[color/.subGC[gNN[p]],1,8] 
  Lam[color/.subGC[gNN[p]],a,b]
  HGVp15[{p,color/.subGC[gNN[p]]/Sqrt[2]}],
  {sum[col6,1,3]sum[col6b,1,3]Lam[col1,col6,col6b],
  {col1->col6,col1b->col6n}}}
 }
};
gNN[g_]:=
     5+ToExpression[StringTake[ToString[g],{-1}]];
\end{MIN}
Each entry consists of the following pieces: 
\begin{itemize}
 \item The name of tensor field ({\tt HG015}, {\tt HGp15})
 \item The name of the gauge group for which the re-writing shall take place ({\tt color})
 \item The name which should be used for the vector representation ({\tt HGV015}, {\tt HGVp15})
 \item The substitution rule:
\begin{verbatim}
 HG015[{p_,a_,b_}] :> 
   sum[color/.subGC[gNN[p]],1,8] Lam[color/.subGC[gNN[p]],a,b] 
   HGV015[{p,color/.subGC[gNN[p]]/Sqrt[2]}]
\end{verbatim}
Here, {\tt p} is an unique index ({\tt gen1}, {\tt gen2}, {\tt gen3}, {\tt gen4}) counting the fields in each interaction term and {\tt gNN} is a function to shift this index 
by 5. Moreoever, {\tt a}, {\tt b} are the colour indices. Therefore, the above line is interpreted as
\begin{equation}
\Phi^p_{\alpha\beta} \to \sum_{f(p)} \lambda^{f(p)}_{\alpha\beta} \frac{1}{\sqrt{2}} \tilde{\Phi}^p_{f(p)}
\end{equation}
with a function $f$ to rename the indices. 
 \item Finally, one needs to define also the reverse operation, i.e. the relation to re-write the vector into the tensor representation. This is needed to derive the ghost interactions. 
\end{itemize}
\end{enumerate}
\item Once the gauge sector and relation for the fields before EWSB are fixed, the rest of the model file is straightforward and follows the standard \SARAH conventions:
\begin{enumerate}
\item {\bf Lagrangian}: the Lagrangian consists of two parts:
\begin{MIN}
 DEFINITION[GaugeES][LagrangianInput]= {
  {LagHC,{AddHC->True}},
  {LagNoHC,{AddHC->False}}	
 };
\end{MIN}
For the first part, the hermitian conjugate needs to be added ({\tt AddHC->True}). This part involves the fermion interactions as well as $\lambda_4$:
\begin{MIN}
LagHC = -(Y1 FQL.FU.H + Y2 FQL.FU.Phi15
    + Y3 conj[H].FQL.FD + Y4 conj[Phi15].FQL.FD
    + Y5 FU.Chi.Si + \[Mu]/2 Si.Si  
    + lambda4 conj[H].conj[Chi].Phi15.Chi);
\end{MIN}                                                                                                                
All other parts of the Lagrangian are already hermitian and are defined via:                                                                                                                
\begin{MIN}
LagNoHC = - (mH2 conj[H].H + mchi2 conj[Chi].Chi 
 + mPhi2 conj[Phi15].Phi15 
 + lambda1 conj[H].H.conj[Chi].Chi 
 + lambda2 conj[H].H.conj[Phi15].Phi15 
 + lambda3 conj[Chi].Chi.conj[Phi15].Phi15 
 + lambda5 conj[H].conj[Phi15].Phi15.H 
 + lambda6 conj[Chi].Phi15.conj[Phi15].Chi 
 + lambda7 conj[H].H.conj[H].H 
 + lambda8 conj[Chi].Chi.conj[Chi].Chi 
 +  Delta[pat1,pat2] Delta[pat2b,pat3b]*
    Delta[pat3,pat4] Delta[pat4b,pat1b]*
    Delta[lef1,lef2] Delta[lef3,lef4]* 
  lambda9 conj[Phi15].Phi15.conj[Phi15].Phi15 
 + Delta[pat1,pat2] Delta[pat2b,pat1b]*
   Delta[pat3,pat4]Delta[pat4b,pat3b]*
   Delta[lef1,lef2] Delta[lef3,lef4] 
  lambda10 conj[Phi15].Phi15.conj[Phi15].Phi15); 
\end{MIN}
For all terms but $\lambda_9$ and $\lambda_{10}$ the index contraction is unique. For those terms one needs to define the contraction explicitly using Kronecker deltas.  The remaining terms coming with $\lambda_{11}$-$\lambda_{19}$ can be implemented in a similiar fashion.
\item {\bf VEVs}: the VEVs are set via
\begin{MIN}
DEFINITION[EWSB][VEVs]={
(* actual Vevs *)
{H0,{vH0,sqr2},{sigmaH0,I*sqr2},{phiH0,sqr2}},
{Chi0,{vChi,sqr2},{sigmaChi,I*sqr2},{phiChi,sqr2}},
{HS015,{vHS15,sqr2},{sigmaHS15,I*sqr2},
{phiHS15,sqr2}},
{HGV015,{0,0},{sigmaV15,I*sqr2},{phiV15,sqr2}}
};
sqr2=1/Sqrt[2];
\end{MIN}
Although the colour octet doesn't receive a VEV, it's CP even and odd component has a different mass. Therefore, it is also decomposed in real fields. 
\item {\bf Gauge bosons}: the rotations of the gauge bosons are defined via
\begin{MIN}
DEFINITION[EWSB][GaugeSector] = {
  {{VS,VWL[3],VWR},{VP,VZ,VZP},ZZ},
  {{VWL[1],VWL[2]},{VWm,conj[VWm]},ZW},
  {{VX,VY},{VLQ,conj[VLQ]},ZLQ}
};  
\end{MIN}
with the rotation matrices defined in the {\tt parameters.m} file as
\begin{equation}
Z^W = Z^{LQ} = \frac{1}{\sqrt{2}}\left(\begin{array}{cc} 1 & 1 \\ i & -i \end{array}\right)
\end{equation}
For the rotation in the neutral sector no explicit parametrisation for {\tt ZZ} is used.
\item {\bf Matter fields}: the rotations in the matter sector are defined via 
\begin{MIN}
DEFINITION[EWSB][MatterSector]= {
(* Neutral scalars *)
 {{phiH0,phiChi,phiHS15},      {hh,ZH}},
 {{sigmaH0,sigmaChi,sigmaHS15},{Ah,ZA}},

(*Charged & coloured scalars*)
 {{HSp15,Hp},              {Hpm,ZP}},
 {{HX015,conj[HY015],Chiu},{Hc0,ZC0}}, 

(*Fermions*)
  {{{dL},{dR}},{{DL,Vd},{DR,Ud}}},
  {{{uL},{uR}},{{UL,Vu},{UR,Uu}}},
  {{{eL},{eR}},{{EL,Ve},{ER,Ue}}},
  {{vL,vR,Sing},{FV,PMNS}} 
}; 
\end{MIN}
\item {\bf Dirac spinors}: the Weyl spinors are combined to Dirac spinors via
\begin{MIN}     
DEFINITION[EWSB][DiracSpinors]={
 Fd -> {DL,conj[DR]},
 Fe -> {EL,conj[ER]},
 Fu -> {UL,conj[UR]},
 Fv -> {FV,conj[FV]}};
\end{MIN}
\end{enumerate}
\end{enumerate}
Useful relations for the generators are used by \SARAH are
\begin{align}
\Lambda^{8+x}_{y4} = & \Lambda^{8+x}_{4y} = \frac12 \delta_{xy} \\
 \Lambda^{11+x}_{y4} = &  -\Lambda^{11+x}_{4y} = -\frac12 i \delta_{xy} 
\end{align}
for $1 \le x,y \le 3$.

For the structure constants, one can make use of
\begin{align}
F_{a b  (c+8)} &=  F_{a b  (c+11)} =  0 \\
F_{i a b } F_{i c d} &=  f_{i a b} f_{i c d} \\ 
F_{c (8+x) (8+y)} &= F_{c (11+x) (11+y)} = i \frac14 (\lambda^c_{xy} - (\lambda^c_{xy})^*) \\
F_{c (8+x) (11+y)} &= - \frac14 (\lambda^c_{xy} + (\lambda^c_{xy})^*) \\
F_{(8+x) (8+y) 15} &= F_{(8+x) (8+y) (8+z)} = F_{(8+x) (8+y) (11+z)} = F_{(8+x) (11+y) (11+z)} =  \nonumber \\
& =  F_{(11+x) (11+y) (11+x)} = F_{(11+x) (11+y) 15} = 0 \\
F_{(8+x) (11+y) 15} &= \sqrt{\frac23} \delta_{xy} 
\end{align}
for $1 \le a,b,c \le 8$  and $1 \le x,y,z \le 3$.

\bibliographystyle{aapmrev4-2}
%

\renewcommand{\bibname}{Bibliography}
\bibliography{LQ.bib}
\end{document}